\documentclass[a4paper,10pt]{article}

\usepackage[latin1]{inputenc}
\usepackage{amsmath}
\usepackage{amsfonts}
\usepackage{amssymb}
\begin{document}

\begin{titlepage}

\topmargin=3.5cm

\textwidth=13.5cm

\centerline{\Large \bf Equivariant Symplectic Geometry of  }

\vspace{0.4cm}

\centerline{\Large \bf  Gauge Fixing in Yang-Mills Theory}

\vspace{1.0cm}

\centerline{\large \bf Levent Akant\footnote{E-mail:
akant@gursey.gov.tr}}

\vspace{0.5cm}

\centerline{ \textit{Feza Gursey Institute}}

\centerline{\textit{Emek Mahallesi, Rasathane Yolu No.68 }}
\centerline{\textit{Cengelkoy, Istanbul, Turkey}}

\vspace{1.5cm}

\paragraph{Abstract:} The Faddeev-Popov gauge fixing in Yang-Mills theory is interpreted as equivariant localization. It is shown that the Faddeev-Popov procedure amounts to a construction of a symplectic manifold with a Hamiltonian group action. The BRST cohomology is shown to be equivalent to the equivariant cohomology based on this symplectic manifold with Hamiltonian group action. 
The ghost operator is interpreted as a (pre)symplectic form and the gauge condition as the moment map corresponding to the Hamiltonian group action. This results in the identification of the gauge fixing action as a closed equivariant form, the sum of an equivariant symplectic form and a certain closed equivariant 4-form which ensures convergence. An almost complex structure compatible with the symplectic form is constructed. The equivariant localization principle is used to localize the path integrals onto the gauge slice. The Gribov problem is also discussed in the context of equivariant localization principle. As a simple illustration of the methods developed in the paper, the partition function of N=2 supersymmetric quantum mechanics is calculated by equivariant localization 
\end{titlepage}

\baselineskip=0.6cm

\abovedisplayskip=0.4cm

\belowdisplayskip=0.4cm

\abovedisplayshortskip=0.3cm

\belowdisplayshortskip=0.3cm

\jot=0.35cm

\section{Introduction}

There are certain field theories with BRST-like symmetries for which
the path integrals localize on small subsets of the space of all
field configurations. Yang-Mills theory on a Riemann surface
\cite{witten1}, Chern-Simons theory on a Seifert manifold
\cite{witten2}, and the so-called $G/G$ models \cite{blau1} provide
examples where this phenomenon takes place. It is well known that
the reason for this localization is that the BRST-like symmetry and
the field content of these models form a differential complex for
equivariant cohomology, and that the action functionals are closed
forms in that differential complex. Localization follows from basic
properties of integration of equivariant forms \cite{witten1,
witten2, KJ}. The aim of this paper is to interpret the
Faddeev-Popov gauge fixing procedure \cite{fp, dw} (see also \cite{ft, th}) as equivariant localization.
Such a reformulation improves our understanding of the BRST symmetry \cite{brs}
and completes its geometric interpretation \cite{RB, thierry} by identifying the antighost and the Lautrup-Nakanishi auxiliary field \cite{lautrup} as geometric objects that arise in the Cartan model of equivariant cohomology. The role of
Gribov horizons \cite{gribov} in equivariant localization will also
be discussed. A discussion in the framework of Batalin-Vilkovisky \cite{bv} formalism will not be attempted at this stage (see \cite{ners, sch}). That will be the subject matter of a future work. 
The relation between BRST cohomology and equivariant cohomology is well known in topological gauge theories \cite{Kalkman, kanno}. Here we will consider the physical gauge theories.

The equivariant cohomology can be thought of as the generalization of the de Rham theory
to manifolds with group actions, where the usual exterior derivative is replaced by the so-called Cartan derivative. If a compact group $G$ acts freely on a manifold $M$ then the corresponding equivariant cohomology is equivalent
to the de Rham cohomology of the quotient manifold $M/G$. If the action is not free then the quotient space $M/G$
is not a smooth manifold and one cannot define the de Rham cohomology of $M/G$. Altough the de Rham cohomology of
the quotient space is not well defined when the group action is not free, the equivariant cohomology still makes
sense (it is just not equivalent to de Rham cohomology of a finite dimensional smooth manifold) \cite{bott, GS}.
It is in this respect that the equivariant cohomology is to be thought of as the generalization of de Rham theory.
In the present paper we will assume the manifold $M$ to be a symplectic manifold and the group action to be
symplectic (canonical) and Hamiltonian \cite{atiyah-bott}. A very interesting property of closed equivariant forms
on such manifolds is that their integrals
localize on the critical points of a certain function which is related to the moment map of the group action.
Equivariant localization principle for abelian group actions was discovered by Duistermaat and Heckman \cite{DH};
while the nonabelian case was developed by Witten in \cite{witten1}, and Kirwan and Jeffrey in \cite{KJ} (for a short review see also \cite{witten3}). Some useful reviews of equivariant cohomology are \cite{moore, szabo, libine}

In this paper we will work with linear covariant gauges for which the gauge fixing action reads
\begin{equation}
S_{gf}=\int d^{d}x\; \mathrm{tr}\left[ \frac{1}{2}\epsilon\, b^{2}-ib\,\partial\cdot A-\overline{c}\,\partial\cdot\nabla c\right]
\end{equation}
where $\nabla_{\mu}=\partial_{\mu}+\mathrm{ad}( A_{\mu})$ is the covariant derivative in the adjoint representation. Our analysis seems to be valid for more general gauge fixing conditions as well.

We will see that the gauge fixing procedure involves the construction of a symplectic manifold with a symplectic and Hamiltonian group action. It is precisely this geometric construction that will allow us to see that the nonminimal BRST complex (which, on top of the gauge field and the ghost, contains also the anti-ghost and the Lautrup-Nakanishi auxiliary field) is in fact a Cartan model of equivariant cohomology. More precisely, the symplectic manifold in question will be a vector bundle over a gauge orbit and the group action will be given by fiber translations.
Ghosts will be interpreted as fundamental 1-forms corresponding to the action of the gauge group \cite{RB} (see also \cite{ht, Azcarraga}), whereas the anti-ghosts will be identified with fundamental 1-forms of fiber translations. The equivariant cohomology based on the fiber translations will be constructed. This cohomology will have the same generators as the BRST complex and, moreover, the action of the Cartan derivative on the generators will be shown to be the same as the action of the BRST operator. The only difference between the two cohomologies will be the gradings assigned to the generators. Altough our group of fiber translations is abelian we will use the language and methods of Witten's nonabelian localization principle, which are more natural for the problem.
The ghost operator will be interpreted as a (pre)symplectic form on our vector bundle and the gauge condition as the moment map corresponding to fiber translations. In fact we will see that the gauge fixing action is an equivariant form; it will be the sum of the so-called
equivariant symplectic form and an equivariant 4-form \footnote{In this respect the gauge fixing action will have the same equivariant structure as the action of Yang-Mills theory on a Riemann surface in first order formalism for which nonabelian localization principle was developed in the first place by Witten in \cite{witten1}.}.
In order to apply equivariant localization principle we will need the pre-symplectic form (ghost operator) to be non-degenerate. It is well known that the ghost operator is degenerate on the so-called Gribov horizons \cite{gribov}. These horizons must be excluded from the region of integration, since they cause over-counting of gauge fields in the path integral. In the standard treatment \cite{gribov} one usually restricts the path integral to the interior of the first Gribov horizon and thus avoids all Gribov horizons. So we will see that the problem encountered in the standard treatment and the one encountered in our analysis based on the use of equivariant localization principle is the same: the degeneracy of the ghost operator. The solution to the problem will be the same: we will, a priori, restrict the region of integration to the interior of the first Gribov horizon. However a consistent use of equivariant localization principle in calculating integrals over a region $R$ (in our case the interior of the first Gribov horizon) with $\partial R \neq \emptyset$ requires, not only the nondegeneracy of the symplectic form on $R$, but also the vanishing of a certain surface term on $\partial R$ (in our case the Gribov horizon). 

The practicality of equivariant localization principle stems from the freedom of choosing certain auxiliary structures used in the localization process. One such structure is an invariant almost complex structure compatible with the symplectic form. We will show that by choosing an appropriate almost complex structure the surface term in question may be set equal to zero on the first Gribov horizon (in fact on all Gribov horizons). This makes possible the use of equivariant localization principle to localize the path integrals for the gauge theory on the gauge slice. We will also see that the exclusion of Gribov horizons is related to the question of independence of the path integral from the gauge fixing parameter $\epsilon$ (gauge independence).

There are certain topological field theories whose actions can be
obtained by gauge fixing the zero action which possesses a very
large transitive abelian gauge symmetry \cite{bs} (also \cite{bs2, bbrt}). Since the starting action
is zero for these theories, the total action after gauge fixing is
the gauge fixing action itself. Therefore it should be possible to
apply our methods to calculate e.g. the partition functions of these
theories via localization. We will apply this strategy to the
partition function of supersymmetric quantum mechanics \cite{witten4, witten5, sh}. In this case we do not have any reason to restrict the path integral inside the first Gribov horizon; we must integrate over the whole space of field histories. We will see that this can possibly lead to gauge dependence of path integrals, if there are
critical points which lie on Gribov horizons. However, we will also see that for field configurations with
periodic boundary conditions there are no such critical points, and this will lead to a gauge independent result.

Most of our calculations will be done in a finite dimensional setting and then the results will be formally generalized to infinite dimensions. It would be interesting to specify infinite dimensional functional spaces where this correspondence becomes rigorous. This will be done elsewhere.
In the present work we will concentrate on the formal algebraic and geometric aspects of the problem. 

Here is the outline of the paper. In Sec. 2 we give an algebraic
introduction to equivariant cohomology and work out an example which
will be used in giving a geometric meaning to anti-ghost and
auxiliary Lautrup-Nakanishi field, constructing the symplectic
manifold in question, and establishing the connection between the
BRST complex and the Cartan model of equivariant cohomology. In Sec.
3 we will consider geometric Cartan model based on a symplectic
manifold with Hamiltonian group actions. We will introduce the
equivariant symplectic form and review the basics of equivariant
integration and nonabelian localization principle. We will end the
section with a discussion of complications that afflict the
localization principle when the equivariant symplectic form has
singularities. In Sec. 4 we will establish the relation between the
BRST cohomology and the Cartan model of equivariant cohomology, from
both algebraic and geometric perspectives. In Sec. 5 we will
generalize the discussion of the previous section to Yang-Mills
theory. We will identify the ghost operator as a symplectic form,
construct the moment map, and interpret the gauge fixing action as a
closed equivariant form. In Sec. 6 we will interpret the 
Faddeev-Popov method as equivariant localization. We will introduce
the complex structure which will be used in localization, and solve
the problems caused by the singularities of the symplectic form. In
Sec. 7 we will illustrate the applicability of our method on a
simple example from topological field theory. Namely, we will
calculate the partition function of SUSY quantum mechanics by
equivariant localization.

\section{Cartan Model}

In this section we will review the basics of algebraic version of equivariant cohomology.
We refer the reader to \cite{GS} for a more detailed account of the subject.

Let $(A,d)$ be a graded commutative differential algebra. Let $G$ be
a group which acts on $(A,d)$ as a group of linear maps that
preserve the grading. Thus if
\begin{equation}
    A=\bigoplus_{i}A^{(i)}
\end{equation}
and $\Phi_{g}$ denotes the action of $g\in G$ then
$\Phi_{g}:A^{(i)}\rightarrow A^{(i)}$. We will denote the Lie
algebra of $G$ by $\mathbf{g}$. $(A,d)$ is called a $G^{*}$ algebra
if there is a graded Lie algebra $\widetilde{\mathbf{g}}$, which is
generated by three types of derivations of $A$: for each $\xi\in
\mathbf{g}$ one has derivations $\pounds_{\xi}$ of degree $0$,
derivations $\iota_{\xi}$ of degree $-1$, and the differential $d$,
of degree $1$, as the generators. The graded commutation relations
of these generators are
\begin{eqnarray}
  \left[\pounds_{\xi}, \pounds_{\eta}\right]&=& \pounds_{[\xi,\eta]} \hspace{1cm} \left[\iota_{\xi}, \iota_{\eta}\right]_{+} = 0 \nonumber\\
  \left[\pounds_{\xi}, \iota_{\eta}\right] &=& \iota_{[\xi,\eta]}\hspace{1.2cm}\left[\iota_{\xi}, d\right]_{+}= \pounds_{\xi}\nonumber \\
  \left[\pounds_{\xi}, d\right] &=& 0\hspace{1.7cm}\left[d, d\right]_{+}= 0.
\end{eqnarray}
If the graded algebra is assumed to have a topological structure
which allows limits to be taken, then one also requires the
compatibility conditions of the actions:
\begin{eqnarray}
\Phi_{g}\pounds_{\xi}\Phi_{g^{-1}}&=&\pounds_{Ad_{g}\xi}\\
\Phi_{g}\iota_{\xi}\Phi_{g^{-1}}&=&\iota_{Ad_{g}\xi}\\
\Phi_{g}d\Phi_{g^{-1}}&=&d.
\end{eqnarray}

A very important example of a $G^{*}$ algebra arises when a
connected Lie group $G$ acts on a manifold $M$. Then $\Omega(M)$ is
a $G^{*}$ algebra. In this case, $\Omega(M)$ is the graded
commutative differential algebra whose grading is simply given by
the exterior degree. The action of $G$ is given by the pull-back
action; while $\tilde{\mathbf{g}}$ is generated by the Lie
derivatives $\pounds_{a}$ along the fundamental vector fields
$V_{a}$ corresponding to the generators $e_{a}$ of $\mathbf{g}$,
contractions $\iota_{a}$ by $V_{a}$, and the exterior derivative
$d$.

In the following we will work at the infinitesimal level, so for us
a $G^{*}$ algebra will be a differential graded commutative algebra
with a $\widetilde{\mathbf{g}}$ action as given above. We will
denote the structure constants of $\mathbf{g}$ in a basis
$\left\{e^{a}\right\}$ by $C^{c}_{ab}$ i.e.
$[e_{a},e_{b}]=C^{c}_{ab}e_{c}$.

Let $\mathcal{S}(\mathbf{g}^{*})$ be the symmetric tensor algebra over $\mathbf{g}^{*}$.
Then for any $G^{*}$ algebra $A$ the space $C(G, A)=\mathcal{S}(\mathbf{g}^{*})\otimes A$ is a graded algebra where the degree of an element of the form
\begin{equation}
    (e^{a_{1}}...e^{a_{r}})\otimes a^{p}
\end{equation}
is given by $p+2r$. Here $a^{p}$ is a homogenous element of $A$ with
degree $p$ and $e^{a}$'s are the basis elements of $\mathbf{g}^{*}$.

 $C(G, A)$ admits a $G$ action
given by
\begin{equation}
    \Psi_{g}=Ad^{*}_{g^{-1}}\otimes \Phi_{g}.
\end{equation}
Here $\Phi_{g}$ is the action of $G$ on $A$ and $Ad^{*}$ is the
coadjoint action of $G$ on $\mathcal{S}(\mathbf{g}^{*})$.
Infinitesimally the action of $\mathbf{g}$ is given by
\begin{equation}
    1\otimes \pounds_{\xi}+\mathcal{C}_{\xi}\otimes 1 \;\;\; \xi \in \mathbf{g}
\end{equation}
Here $\mathcal{C}_{\xi}$ is the coadjoint
action of $\mathbf{g}$ on $\mathcal{S}(\mathbf{g}^{*})$ given by
\begin{equation}
    \mathcal{C}_{\xi}(e^{a_{1}}\ldots e^{a_{r}})=\sum_{n=1}^{r} e^{a_{1}}\ldots
    e^{a_{n-1}}(-\xi^{a}C_{ab}^{a_{n}}e^{b})e^{a_{n+1}}\ldots e^{a_{r}}.
\end{equation}
Note that
\begin{equation}
(e^{a}\mathcal{C}_{a})(e^{a_{1}}\ldots e^{a_{r}})=0.
\end{equation}
The Cartan derivative is defined as
\begin{equation}
    D= 1\otimes d-ie^{a}\otimes \iota_{a}
\end{equation}
Note also that the combination $ e^{a}\otimes \iota_{a}$ is independent of
basis. $D$ is a graded derivation of degree $1$; it increases the
degree of a homogenous element by $1$, and for a homogenous element
$a \in C(G, A)$ of degree $p$ one has
\begin{equation}
    D(ab)=(Da)b+(-1)^{p}a(Db).
\end{equation}
A simple calculation shows that
\begin{equation}
    D^{2}= -ie^{a}\otimes \pounds_{a}.
\end{equation}
Therefore on the space $C_{G}(A)=(C(G, A))^{G}$ of $G$-invariant
elements of $ C(G,A)$ one has
\begin{equation}
    D^{2}=0.
\end{equation}
So, all the ingredients of a cohomology theory are present. The
resulting cohomology is called the equivariant cohomology. The space
$C_{G}(A)$ is called the Cartan model of the equivariant cohomology
of $A$. In what follows we will concentrate on the case
$A=\Omega(M)$. In our local considerations we will take $M$ to be a
vector space $V$. In this case it will be convenient to identify
$\Omega(V)$ with $C^{\infty}(V)\otimes \wedge V^{*}$. Thus
\begin{equation}
C(G,\Omega(V))= \mathcal{S}(\mathbf{g}^{*})\otimes
C^{\infty}(V)\otimes \wedge V^{*} .
\end{equation}
The appropriate graded algebra to use in the field theory context is given by the direct generalization of this finite dimensional model
\begin{eqnarray}
C(b,\pi,\overline{c})=\mathbb{C}[b^{a}]\otimes\mathbb{C}[\pi^{i}]\otimes \wedge [\overline{c}^{i}] \\
a=1,\ldots, \dim G; \ i=1, \ldots, \dim V. \nonumber
\end{eqnarray}
Here $\mathbb{C}[b^{a}]$, $\mathbb{C}[\pi^{i}]$ and $\wedge
[\overline{c}^{i}]$ are the sets of formal power series in the
commuting fields $b^{a}$, $\pi^{i}$ and the anti-commuting field
$\overline{c}^{i}$, respectively. The field $b^{a}$ carries a Lie
algebra index and transforms under the coadjoint action of the group
$G$. Notice that the fields $\pi^{i}$ and $\overline{c}^{i}$ have
the same index structure. The gradings of $b$, $\pi$ and
$\overline{c}$ are 2, 0 and 1, respectively.

Let us consider an example of a Cartan model which will turn out to
be crucial for our constructions in the following. Let $G=V$,
considered as translation group, and $M=V$. Then the action of $V$
on $C(V,\Omega(V))=\mathcal{S}(V^{*}) \otimes \Omega(V)$ is given by
\begin{equation}
\Phi_{v}=1\otimes T_{v}
\end{equation}
where $T_{v}$ is a translation in $V$ by a vector $v\in V$. So we have an action of $V$ on $C(V, \Omega(V))$.
The infinitesimal action is given by
\begin{equation}
1\otimes \pounds_{v}.
\end{equation}
The Cartan differential is given by
\begin{equation}
D=1\otimes d-ie^{a}\otimes \iota_{a}.
\end{equation}
The Cartan model is the $V$ (translation) invariant part of $C(V, \Omega(V))=\mathcal{S}(V^{*})\otimes C^{\infty}(V)\otimes \wedge V^{*}$
\begin{equation}
C_{V}(\Omega(V))=\mathcal{S}(V^{*})\otimes \wedge V^{*}.
\end{equation}
In the field theoretical model $C(b,\pi,\overline{c})$ the group action on the generators is given by
\begin{eqnarray}
\Phi_{v}b^{a}&=&0\\
\Phi_{v}\pi^{a}&=&\pi^{a}+v^{a}\\
\Phi_{v}\overline{c}^{a}&=&0.
\end{eqnarray}
Infinitesimally,
\begin{eqnarray}
\pounds_{v}b^{a}&=&0\\
\pounds_{v}\pi^{a}&=&-v^{a}\\
\pounds_{v}\overline{c}^{a}&=&0.
\end{eqnarray}
The action of the Cartan derivative on the generators is given by
\begin{eqnarray}
Db^{a}&=&0\\
D\pi^{a}&=&\overline{c}^{a}\\
D\overline{c}^{a}&=&-ib^{a}.
\end{eqnarray}
Finally, the Cartan model is the translationally invariant part (i.e. the part annihilated by all $\pounds_{v}$'s) of $\mathbb{C}[b^{a}]\otimes\mathbb{C}[\pi^{a}]\otimes \wedge [\overline{c}^{a}]$
  \begin{eqnarray}
C_{G}(b,\overline{c})=\mathbb{C}[b^{a}]\otimes \wedge [\overline{c}^{a}].
\end{eqnarray}

\section{Integration and Localization}
In this section we consider the geometric model $C_{G}(\Omega(M))$
for equivariant cohomology. We will assume  $M$ to be a compact
symplectic manifold without boundary. The symplectic form on $M$
will be denoted by $\omega$. We will assume that the action of the compact
connected group $G$ on $M$ is symplectic (canonical) and
Hamiltonian. We will also assume that the Lie algebra $\mathbf{g}$
of $G$ is semi-simple. One may relax the compactness
condition on $M$ by considering equivariant forms of compact support
or of rapid decrease.

\subsection{Cartan Model on a Symplectic Manifold}
If $\Phi_{g}$ denotes the action of $g \in G$ on $M$ then
\begin{equation}
\Phi_{g}^{*}\omega=\omega.
\end{equation}
Infinitesimally the action of $u \in \mathbf{g}$ is given by the fundamental vector field $V_{u}$.
Thus we have
\begin{equation}
\pounds_{V_{u}}\omega=0.
\end{equation}
Let $\left\lbrace T_{a}\right\rbrace $ be a basis for $\mathbf{g}$, then we will denote $V_{T_{a}}$ simply by $V_{a}$, and the corresponding Lie derivative and contraction by $\pounds_{a}$ and $\iota_{a}$, respectively.

The action of $G$ is assumed to be Hamiltonian. This means that for each $u \in \mathbf{g}$ there will be a map $\mu_{u}\in C^{\infty}(M)$, linear in $u$, and satisfying
\begin{equation}
d\mu_{a}=-\iota_{a}\omega
\end{equation}
\begin{equation}
\left\lbrace \mu_{a}, \mu_{b} \right\rbrace=C_{ab}^{c}\mu_{c}.
\end{equation}
Here $C_{ab}^{c}$ are the structure constants of $\mathbf{g}$,
$\mu_{a}=\mu_{T_{a}}$, and $\left\lbrace \;\;,\;\; \right\rbrace$ is
the Poisson bracket induced by the symplectic form $\omega$.

Cartan model of equivariant cohomology $H_{G}(M)$ is the cohomology based on the graded differential algebra
\begin{eqnarray}
C_{G}(\Omega(M))&=&(\mathcal{S}(\mathbf{g^{*}})\otimes \Omega(M))^{G}\\
D&=&d-ib^{a}\iota_{a}.
\end{eqnarray}
A generic equivariant form in $C_{G}(\Omega(M))$ can be written as
\begin{equation}
\sum_{k}\; \alpha^{k}P_{k}(b)
\end{equation}
where $\alpha^{k}$ is a $k$ form in $\Omega(M)$ and $P_{k}(b)$ is a polynomial in $b$ (i.e. an element of $\mathcal{S}(\mathbf{g^{*}})$ ).

A result of fundamental importance in equivariant cohomology is:
assuming the action of $G$ on $M$ is free, the equivariant
cohomology is equivalent to the de Rham cohomology of the quotient
manifold $M/G$ \cite{GS}
\begin{equation}
H_{G}(M)=H_{dR}(M/G).
\end{equation}
In this sense, the equivariant cohomology may be regarded as the generalization of the de Rham cohomology of quotient manifolds to the case of non-free group actions, for which $M/G$ is not a smooth manifold.

Let us examine some examples of equivariant forms. An important example is the equivariant symplectic form \cite{GS}
\begin{equation}
\overline{\omega}=\omega-ib^{a}\mu_{a}.
\end{equation}
It is easy to check that this is a closed equivariant 2-form. For example $G$-invariance can be seen as follows:
\begin{equation}
\pounds_{a}\overline{\omega}=\pounds_{a}\omega-ib^{b}\pounds_{a}\mu_{b}+if_{ab}^{c}b^{b}\mu_{c}=0
\end{equation}
since $\pounds_{a}\omega=0$ by the hypothesis that the action is symplectic, and
\begin{equation}
-ib^{b}\pounds_{a}\mu_{b}+if_{ab}^{c}b^{b}\mu_{c}=-ib^{b}\left(\left\lbrace \mu_{a}, \mu_{b}   \right\rbrace-f_{ab}^{c}\mu_{c} \right)=0
\end{equation}
by the definition of the Poisson bracket and the hypothesis that the action is Hamiltonian. Closedness follows from a similar calculation:
\begin{equation}
D\overline{\omega}=D(\omega-ib^{a}\mu_{a})=d\omega-ib^{a}\iota_{a}\omega-ib^{a}d\mu_{a}=0.
\end{equation}

Conversely suppose $M$ is a manifold on which a group $G$ acts. Let $\overline{\omega}$ be an equivariant two form, which is necessarily of the form
\begin{equation}
\overline{\omega}=\omega-ib^{a}\mu_{a}.
\end{equation}
If $\overline{\omega}$ is closed then we have
\begin{eqnarray}
D\overline{\omega}=d\omega-ib^{a}i_{a}\omega-ib^{a}d\mu_{a}=0.
\end{eqnarray}
This implies
\begin{eqnarray}
d\omega=0,\;\;\;\;\iota_{a}\omega=-d\mu_{a}.
\end{eqnarray}
Moreover the $G$-invariance of $\overline{\omega}$ implies
\begin{eqnarray}
\pounds_{a}\omega=0,\;\;\;\;\pounds_{a}\mu_{b}=C_{ab}^{c}\mu_{c}.
\end{eqnarray}
 If $\omega$ is assumed to be non-degenerate then these observations imply that
 $\omega$ is a symplectic form and $G$ has a symplectic and Hamiltonian action on $M$.

Another example of an equivariant form is the 4-form
\begin{equation}
b\cdot b\equiv(b,b)
\end{equation}
where $(\;\;,\;\;)$ is the Cartan-Killing metric on $\mathbf{g}^{*}$. The $G$-invariance follows from the $Ad$-invariance of the Cartan-Killing metric. It is also easy to see that $(b,b)$ is a closed equivariant form.  This 4-form and the equivariant symplectic form will play important roles in the interpretation of the gauge fixing action as a closed equivariant form.

\subsection{Equivariant Localization}
The integral of an equivariant form is defined as \cite{witten1, witten2}
\begin{equation}
    \int \; \sum_{k}\; \alpha^{k}P_{k}(b)=\int db^{1}\ldots
    db^{N}e^{-\frac{\epsilon}{2} b\cdot b}P_{k}(b) \int_{\mathcal{M}}\alpha^{k}.
\end{equation}
Note that the exponential of the 4-form $b\cdot b$ is a convergence factor which regulates the
integration over $b$'s.

A very important property of this definition is the vanishing of the integral of an exact equivariant form
\begin{equation}
\int D \gamma=\int db^{1}\ldots
    db^{N}e^{-\frac{\epsilon}{2} b\cdot b}P_{k}(b) \int_{\mathcal{M}}d\alpha^{k}.
\end{equation}
which vanishes, by Stoke's theorem, if $M$ is compact and without boundary; or if $\alpha^{k}$ is of compact support or of rapid decrease.
In particular, for any closed equivariant $r$-form $\gamma$ and any equivariant 1-form $\lambda$, the form
\begin{eqnarray}
\gamma(e^{tD\lambda}-1)
\end{eqnarray}
is exact
\begin{eqnarray}
\gamma(e^{tD\lambda}-1)&=&\gamma\left(tD\lambda+\frac{t^{2}}{2!}D\lambda D\lambda+\ldots \right)\\
&=&\gamma\;D\left(t\lambda+\frac{t^{2}}{2!}\lambda D\lambda+\ldots \right)\\
&=&D\left[(-1)^{r}\gamma\left(t\lambda+\frac{t^{2}}{2!}\lambda D\lambda+\ldots \right) \right].
\end{eqnarray}
Consequently,
\begin{equation}
    \int \gamma=\int \gamma\;e^{tD\lambda}.
\end{equation}
This formula is the basis of Witten's nonabelian localization principle \cite{witten1}.
The evaluation of the integral on the right hand side in the large $t$ limit localizes the integral on the critical points of $b^{a}\iota_{a}\lambda$
\begin{eqnarray}
d\;\left( b^{a}\iota_{a}\lambda\right) =0.
\end{eqnarray}
More explicitly, we have two sets of equations:
\begin{eqnarray}{\label{sp1}}
b^{a}\;d\left( \iota_{a}\lambda\right) =0,
\end{eqnarray}
\begin{eqnarray}{\label{sp2}}
\iota_{a}\lambda&=&0.
\end{eqnarray}
These equations can be simplified if $\lambda$ is chosen in a certain way \cite{witten1}. Let $J$ be a $G$-invariant, almost complex structure compatible with the symplectic form $\omega$ (i.e. $\omega(JX,JY)=\omega(X,Y)$) and choose
\begin{equation}
\lambda=JdI.
\end{equation}
Here
\begin{equation}
I=\mu\cdot \mu
\end{equation}
where $\mu=\mu_{a}T^{a}$.
The compatibility of $J$ with $\omega$ implies the metric $g$ defined by
\begin{equation}
g(X,Y)=\omega(X,JY)
\end{equation}
is positive definite. With this choice of $\lambda$, and under the assumption made about $J$, one can arrive at the following two conclusions whose proofs rely heavily on the positive definiteness of $g$ \cite{witten1}:
\begin{eqnarray}{\label{c1}}
\iota_{a}\lambda=0 \Leftrightarrow dI=\sum_{a}\mu_{a}d\mu_{a}=0,
\end{eqnarray}
and that under the assumption of free action of $G$ on $\mu^{-1}(0)$
\begin{equation}{\label{c2}}
b^{a}\left.d\left(
\iota_{a}\lambda\right)\right|_{q}=0\Leftrightarrow
b^{a}=0\;\;\;\;for\;\;\;\;q\in \mu^{-1}(0).
\end{equation}

The first conclusion implies two types of critical points: ordinary
critical points of $I$ which satisfy $\mu_{a}=0$ for all $a$; and
higher critical points for which $dI=0$ but $I\neq 0$. We will
denote the set of ordinary critical points by $\mu^{-1}(0)$. The
second conclusion implies the compactness of the ordinary critical
point set in $\mathbf{g}\times \mu^{-1}(0)$ along the $\mathbf{g}$
direction. In particular, the convergence factor
$e^{-\frac{\epsilon}{2} b\cdot b}$ is not needed to make sense of
the integral. In fact, it can be argued \cite{witten1} that on
$\mu^{-1}(0)$ the equivariant integral is independent of the
convergence factor (i.e. independent of $\epsilon$). However, for
higher critical points dependence on $\epsilon$ cannot be avoided.
Moreover, singularities of $g$ may very well lead to $\epsilon$
dependence of the equivariant integral, even for ordinary critical
points. Since $g$ is derived from the symplectic form $\omega$, the
singularities of the latter invalidate conditions (\ref{c1}) and
(\ref{c2}). Depending on the integrand, this may give rise to
$\epsilon$ dependence in the equivariant integral. As we will see below, $\epsilon$ dependence is something we do not want.
\subsection{Gauge Theory}
In our discussion of gauge fixing as equivariant localization
\begin{equation}
    -\int d^{d}x\; \mathrm{tr}\left[\overline{c}\,\partial\cdot\nabla c\right]
\end{equation}
will be interpreted as a (pre)symplectic form and the part of the gauge
fixing action given by
\begin{equation}
    -\int d^{d}x\; \mathrm{tr}\left[ -ib\,\partial\cdot A-\overline{c}\,\partial\cdot\nabla c\right]
\end{equation}
will become an equivariant symplectic form. The Boltzmann factor
corresponding to the remaining part of $S_{gf}$,
\begin{eqnarray}
\exp\left[ -\frac{\epsilon}{2}\int d^{n}x\, b\cdot b\right],
\end{eqnarray}
which is the source of gauge dependence, will be interpreted as the
equivariant convergence factor. A detailed discussion of algebraic and geometric constructions which are responsible for these identifications will be given in the next section. For now we will be contented with some remarks concerning the gauge dependence and degeneracies of the (pre)symplectic form 

The possibility of omitting the convergence factor is closely related to the stability of the BRST
symmetry of the gauge fixed action\footnote{Differentiation with
respect to $\epsilon$ inserts a BRST exact term to the path
integral. Consequently, non-vanishing of the $\epsilon$ derivative
implies non-vanishing of the expectation value of a BRST exact term
and the BRST symmetry is broken.}. We will investigate in Sec. 6 the
conditions under which the convergence factor may be omitted in the
gauge fixed path integral. However, notice once more that the
validity of (\ref{c1}) and (\ref{c2}) is guaranteed only when the
metric $g$ is positive definite.  In our discussion of gauge fixing
we will take the symplectic form $\omega$ to be the ghost Lagrangian
and both $J$ and $g$ will be given in terms of the ghost operator.
Thus, in order to ensure the validity of (\ref{c1}) and (\ref{c2}),
we will have to restrict the region of integration to a subset where
$\omega$ is non-degenerate. However, with the restriction of the
equivariant integral to a subregion $R\subset M$ with $\partial
R\neq \emptyset$ we must question the validity of the equivariant
localization principle, since the difference
\begin{equation}
\Delta_{\gamma}(R)=\int_{R} \gamma(1-e^{tD\lambda}),
\end{equation}
which, in general, reduces to a surface integral over $\partial R$, may not vanish.
One can safely use the equivariant localization principle in the region $R$ only when $\Delta_{\gamma}(R)=0$.

Summarizing, in all our applications of equivariant localization we
will require the validity of the conditions (\ref{c1}) and
(\ref{c2}). If they are not satisfied at certain singular points
(e.g. points where $g$ is singular or points where $\omega$ is
degenerate), then we will avoid them by restricting the
equivariant integral to a region which contains no singularity. But
in such a region equivariant localization principle may no longer be
applicable due to a non-vanishing surface term. In such cases the
validity of localization principle should be checked by showing that
$\Delta_{\gamma}(R)=0$ on the restricted region.

By a Gribov horizon we will mean a connected region in the space of
all gauge fields where the ghost operator is degenerate. As was
argued by Gribov in \cite{gribov}, these horizons may be thought of
as a collection of bounding surfaces, with the property that the
$k^{th}$ horizon lies entirely in the region bounded by $(k+1)^{st}$
horizon. Moreover, on the first horizon there is only one zero mode
of the ghost operator, on the second there are two, on the third
there are three, and so on. In a region lying between two
consecutive horizons the ghost operator is non-degenerate. As was
observed by Gribov \cite{gribov}, these horizons should be avoided
in path integrals. Gribov horizons contain gauge fixed field
configurations in whose neighborhoods there are other gauge fields
which obey the same gauge fixing condition but differ from each
other by infinitesimal gauge transformations (i.e. they lie on the
same gauge orbit); clearly such configurations lead to over-counting
in the path integral and therefore should be eliminated altogether.
The standard procedure is to restrict the path integral to the
interior of the first Gribov horizon. But on this region $\omega$ is
non-degenerate and the conditions (\ref{c1}) and (\ref{c2}) are
satisfied. We will see that when the integral is restricted inside the first Gribov horizon,
$\Delta_{\gamma}(R)=0$ for $\gamma=e^{\overline{\omega}}$ and for an appropriate choice of the almost complex structure $J$.

\section{BRST Complex as a Cartan Model}
\subsection{Algebraic Model}
In this section we will construct an algebraic model for equivariant cohomology and relate it to the BRST cohomology. Geometrically oriented readers may safely skip this section and proceed to the next one where a geometric version of the same construction is discussed.
Let $ B(K,\mathcal{O})=\wedge \mathbf{k^{*}}\otimes
C^{\infty}(\mathcal{O})$ be the minimal sector (no anti-ghosts and
no auxiliary fields) of the BRST complex. Here we are assuming that
$K$ is a connected group which acts freely on a manifold $N$ and
that $\mathcal{O}$ is a $K$-orbit in $N$. $\mathbf{k}$ denotes the
Lie algebra of $K$ and $C^{\infty}(\mathcal{O})$ is the set of
smooth functions on $\mathcal{O}$. In field theory context we will
use the generalization of the algebraic model
\begin{eqnarray}
 B(K, \mathcal{O})=\mathbb{C}[q^{a}]\otimes
\wedge[c^{b}] \\
a,b=1, \ldots, \dim \mathcal{O}=\dim K \nonumber
\end{eqnarray}
where $N$ will be identified with the space of gauge connections,
$\mathcal{O}$ with a gauge orbit and $K$ with the gauge group. Here
$q^{a}$ are the coordinates on $\mathcal{O}$.

The action of the BRST differential $s$ on
the generators is given by
\begin{eqnarray}
  sf(q) &=& c^{a}\pounds_{e_{a}}f(q),\;\;\;\;f\in C^{\infty}(\mathcal{O}) \\
  sc^{a} &=&-\frac{1}{2}c^{b}c^{c}f^{a}_{bc}
\end{eqnarray}
Here $\pounds_{e_{a}}$ is the Lie derivative in the direction of the fundamental vector field $e_{a}$ corresponding
to the generator $T_{a}$ of $\mathbf{k}$, and $\left[T_{a},T_{b} \right]=f^{c}_{ab}T_{c}$.

Now let us consider the tensor product of graded algebras
\begin{equation}
C(G, A)\otimes B(K, \mathcal{O})=\mathcal{S}(\mathbf{g}^{*})\otimes A \otimes
B(K, \mathcal{O}).
\end{equation}
Write
\begin{eqnarray}
A&=&\bigoplus_{i}A_{i} \\
B(K, \mathcal{O})&=&\bigoplus_{j}B_{j}
\end{eqnarray}
where $A_{i}$ is the space of homogenous elements of $A$ with degree $i$, and likewise $B_{j}$ are the elements
of $B$ with ghost number $j$. Then $\mathcal{A}=A \otimes B(K, \mathcal{O})$ is a graded algebra
\begin{equation}
\mathcal{A}=\bigoplus_{k}\mathcal{A}_{k}
\end{equation}
\begin{equation}
\mathcal{A}_{k}=\mathrm{span}\left\lbrace a^{i}\otimes b^{j}:a^{i}\in A_{i},\  b^{j}\in B_{j}, \ i+j=k \right\rbrace
\end{equation}
We define the product on $\mathcal{A}$ as
\begin{equation}
(a^{i}\otimes b^{j})(a^{r}\otimes b^{s})=(-1)^{jr}(a^{i}c^{r}\otimes b^{j}b^{s}).
\end{equation}
From this definition it follows that (i)
$\mathcal{A}_{k}\mathcal{A}_{l}\subset \mathcal{A}_{k+l}$ and (ii)
$\mathcal{A}$ is super-commutative. Thus $\mathcal{A}$ is a graded
commutative algebra.
 \newtheorem{sub}{Proposition}[section]
\begin{sub}
  $\mathcal{A}$ is a $G^{*}$ algebra.
\end{sub}
\textit{Proof:}
We already remarked that $\mathcal{A}$ is a graded commutative algebra. We will define the action of $G$ on
$\mathcal{A}=A \otimes B(K, \mathcal{O})$ as
\begin{equation}
\Psi_{g}= \Phi_{g} \otimes 1.
\end{equation}
This is clearly a well defined action.
The associated Lie superalgebra is generated by
\begin{eqnarray}
 &\pounds_{\xi}\otimes 1&\\
 &\iota_{\xi}\otimes 1&\\
 &d\otimes 1+(-1)^{F} \otimes s.&
\end{eqnarray}
Here $\pounds_{\xi}$ denotes the infinitesimal action of $\xi\in \mathbf{g}$ on $A$, and $F$ is the number
operator for the grading of $A$. The commutation relations of $(-1)^{F}$ with the generators are
\begin{equation}
\left[(-1)^{F}, \pounds_{\xi} \right]= \left[(-1)^{F}, \iota_{\xi} \right]_{+}=\left[(-1)^{F}, d \right]_{+}=0
\end{equation}
Consequently one gets
\begin{eqnarray}
\left[ \pounds_{\xi}\otimes 1, \pounds_{c}\otimes 1\right] &=&\pounds_{\left[ \xi,c\right] }\otimes 1\\
\left[ \pounds_{\xi}\otimes 1,\iota_{c}\otimes 1 \right] &=&\iota_{\left[ \xi,c\right] }\otimes 1\\
\left[ \pounds_{\xi}\otimes 1,d\otimes 1+(-1)^{F}\otimes s \right] &=&0\\
\left[\iota_{\xi}\otimes 1,\iota_{c}\otimes 1 \right]_{+} &=&0\\
\left[\iota_{\xi}\otimes 1,d\otimes 1+(-1)^{F} \otimes s  \right]_{+} &=&\pounds_{\xi}\otimes 1\\
\left[d\otimes 1+(-1)^{F} \otimes s ,d\otimes 1+(-1)^{F} \otimes s  \right]_{+} &=&0.
\end{eqnarray}
Thus we conclude that $\mathcal{A}= A\otimes  B(K,\mathcal{O})$ is a $G^{*}$ algebra.

As a corollary to this proposition we have

\begin{sub}
  $C(G,A)\otimes B(K,\mathcal{O})=C(G,\mathcal{A})$.
\end{sub}
The Cartan derivative on $C(G,\mathcal{A})$
is defined according to the standard construction as
\begin{equation}
D=1\otimes d\otimes 1+1\otimes (-1)^{F} \otimes s-ib^{a}\otimes \iota_{a}\otimes 1
\end{equation}
The Cartan model is given by
\begin{equation}
C_{G}(\mathcal{A})=(\mathcal{S}(\mathbf{g}^{*})\otimes A\otimes
B(K,\mathcal{O}))^{G}=(\mathcal{S}(\mathbf{g}^{*})\otimes A)^{G}\otimes B(K,\mathcal{O})
\end{equation}
since the action of $G$ on $ B(K,\mathcal{O})$ is trivial. Moreover if
$\mathbf{g}$ is abelian then
\begin{equation}
C_{G}(\mathcal{A})=\mathcal{S}(\mathbf{g}^{*})\otimes A^{G}\otimes
B(K,\mathcal{O})
\end{equation}

Now let us consider the special case $C(G,A)=C(\mathbf{k},\Omega(\mathbf{k}))$ where, as we did in the example
at the end of Sec.2, we take $\mathbf{k}$ as an abelian group acting on itself by translations. In this case
\begin{equation}
C(\mathbf{k},\mathcal{A})=\mathcal{S}(\mathbf{k^{*}})\otimes (
C^{\infty}(\mathbf{k})\otimes \wedge \mathbf{k^{*}})\otimes ( C^{\infty}(\mathcal{O})\otimes\wedge
\mathbf{k^{*}} ).
\end{equation}
The Cartan derivative on the generators of
$C(\mathbf{k},\mathcal{A})$ is
\begin{eqnarray}
Db^{a}&=&0\\
D\pi^{a}&=&\overline{c}^{a}\\
D\overline{c}^{a}&=&-ib^{a}\\
Dc^{a}&=&-\frac{1}{2}c^{b}c^{c}C^{a}_{bc}\\
Df(q)&=&c^{b}\pounds_{e_{b}}f(q),\;\;\;\; f\in C^{\infty}(\mathcal{O})
\end{eqnarray}
In the corresponding Cartan model we have $A^{G}= \wedge \mathbf{k^{*}}
$. So the Cartan differential on the generators of
$C_{G}(\mathcal{A})=\mathcal{S}(\mathbf{k^{*}})\otimes \wedge \mathbf{k^{*}}
\otimes (\wedge \mathbf{k^{*}}\otimes C^{\infty}(\mathcal{O})) $  is
\begin{eqnarray}
Db^{a}&=&0\\
D\overline{c}^{a}&=&-ib^{a}\\
Dc^{a}&=&-\frac{1}{2}c^{b}c^{c}C^{a}_{bc}\\
Df(q)&=&c^{b}\pounds_{e_{b}}f(q).
\end{eqnarray}
But this is nothing but the nonminimal sector of the BRST complex,
except the fact that the grading is different. The comparison
between the two gradings is given in the table below where $F$ is
the number operator in the Cartan model and $\Gamma$ is the ghost
number operator in the BRST complex.
\begin{eqnarray}
\begin{tabular}{l|lllr}
  & $b$ & $\overline{c}^{a}$ & $c^{a}$  & $x$ \\ \hline
F & 2 & 1 & 1 & 0 \\
$\Gamma$ & 0 & -1 & 1 & 0
\end{tabular}
\end{eqnarray}
The fact that $D$ is a differential in both complexes follows from the simple observation that $F$ is even
if and only if $\Gamma$ is even.
\subsection{Geometric Model}
Now we will construct a geometric version of our algebraic model. We know that in a geometric setting the Cartan derivative acting on the coordinate functions gives the exterior differentials of the latter.
So from
\begin{equation}
 D\pi=\overline{c}
\end{equation}
we deduce that $\overline{c}$ should be interpreted as the exterior differential of $\pi$.
Similarly,
\begin{equation}
Df(q)=df(q)=c^{a}\pounds_{e_{a}}f(q)
\end{equation}
suggests the interpretation of $c$'s as the 1-forms dual to the fundamental vector fields corresponding to the action of the gauge group $K$. This is in accordance with the standard geometric interpretation of ghost fields \cite{RB} (also \cite{ht}, \cite{Azcarraga}). So the underlying manifold can be taken to be  $\mathcal{O}_{\alpha}\times \mathbf{k}$ where $\mathcal{O}_{\alpha}$ is an orbit of the gauge group
$K$. If we think of this manifold as a trivial vector bundle over $\mathcal{O}_{\alpha}$ with a typical fiber given by the gauge Lie algebra $\mathbf{k}$ then group action has the nice interpretation of fiber translations.
Notice that the dimensions of the base manifold and the fiber are the same. Consequently our vector bundle is even dimensional. Now our aim is to turn a subset of this vector bundle into a symplectic manifold on which the action of fiber translations is symplectic and Hamiltonian. In the next section we will show that Faddeev-Popov gauge fixing in Yang-Mills theory involves a field theoretic version of this construction. 

Explicitly our geometric construction goes as follows. We have a manifold $Q$ and a group $K$ with a free action on $Q$. Let $\mathbf{k}$ be the Lie algebra of $K$ and assume $\mathrm{dim}\mathbf{k} =n$. Let $f^{c}_{ab}$ be the structure constants of $\mathbf{k}$.
In the field theory context $Q$ and $K$ will be the space of connections and the gauge group, respectively. Let us label the orbits $\mathcal{O}_{\alpha}$ of the $K$-action by the index $\alpha$.
Since the action is free, the fundamental vector fields $e_{a}$ form
a global frame for the tangent bundle $T\mathcal{O}_{\alpha}$. Similarly the dual 1-forms $c^{a}$ form a global frame for the cotangent bundle.

Now let us consider $M=Q \times \mathbf{k}$. The coordinates in $Q$ and $\mathbf{k}$ will be denoted by
$x$ and $\pi$ respectively. The direct product $K \times \mathbf{k}$ has a natural action on
$M$. Here we think of $\mathbf{k}$ as an additive group acting on itself by translations. It is this action, and not the action of $K$ (gauge group), that will be used in the construction of the equivariant cohomology.
The fundamental vector field corresponding to the action of $u \in \mathbf{k}$ on $M$ is
\begin{equation}
V_{u}=u^{a}\frac{\partial}{\partial  \pi^{a}}=u^{a}V_{a}
\end{equation}
where $V_{a}=\frac{\partial}{\partial \pi^{a}}$ and we will use the notation $\overline{c}^{a}\equiv d\pi^{a}$ for the dual basis. Moreover the Lie derivative along $V_{a}$ will be denoted by $\pounds_{a}$ and the contraction $\iota_{V_{a}}$ by $\iota_{a}$.

Under the action of $K$, $M$ is fibered into orbits of the form $\mathcal{U}_{\alpha}=\mathcal{O}_{\alpha} \times \mathbf{k}$. Let
$\omega_{ab}$ be an $n \times n$ matrix valued function on $Q$ which satisfies
\begin{eqnarray}{\label{q}}
\pounds_{e_{c}}\omega_{ab}-\pounds_{e_{b}}\omega_{ac}-f_{cb}^{d}\omega_{ad}=0.
\end{eqnarray}
The meaning of this condition will become clear below. Denote the
restriction of $\omega_{ab}$ to $\mathcal{O}_{\alpha}$ by
$\omega^{(\alpha)}_{ab}$. We will define Gribov horizons as the
connected  components of the solution set of the equation
$\det\,\omega_{ab}=0$. More precisely, the $i^{th}$ Gribov horizon
will be defined as the locus where $\omega_{ab}$ has precisely $k$
vanishing eigenvalues. The $i^{th}$ Gribov horizon will be denoted
by $\ell_{i-1}$. Furthermore, we will assume that Gribov horizons
are boundaries and that $\ell_{i-1}$ lies entirely in the region
bounded by $\ell_{i}$. The region bounded by $\ell_{i-1}$ and
$\ell_{i}$ will be denoted by $C_{i-1}$. In particular the
region bounded by the first Gribov horizon $\ell_{1}$ will be
denoted by $C_{0}$.  Each $\ell_{i-1}$ is naturally
embedded in $M$ as $\ell_{i-1}\times \mathbf{k}$. We will denote the intersections
$C_{i}\cap \mathcal{O}_{\alpha}$ and $\ell_{i}\cap\mathcal{O}_{\alpha}$ by $C_{i}^{(\alpha)}$ and $\ell_{i}^{(\alpha)}$, respectively. Thus, in particular, $\ell_{1}^{(\alpha)}$ is the surface where
the first Gribov horizon intersects the gauge orbit $\alpha$, and $C_{0}^{(\alpha)}$ is the portion of the
gauge orbit bounded by $\ell_{1}^{(\alpha)}$.
The 2-form
\begin{equation}
\omega=\omega_{ab}(x) \overline{c}^{a}\wedge c^{b}\in \Omega^{2}(M)
\end{equation}
is clearly degenerate at each point of $M$. Consider the restriction of $\omega$ to $\mathcal{U}_{\alpha}$
\begin{equation}
\left. \omega\right\vert_{\mathcal{U}_{\alpha}}=\omega^{(\alpha)}_{ab}(q)\overline{c}^{a}\wedge c^{b}
\end{equation}
where $q$'s are the coordinates in $\mathcal{O}_{\alpha}$. This restricted 2-form is degenerate only on
$\ell_{i}^{(\alpha)}\times \mathbf{k}\subset \mathcal{U}_{\alpha}$.
We will also refer to $\ell_{i}^{(\alpha)}\times \mathbf{k}$'s as
Gribov horizons. In the rest of this section we will assume
$\omega^{(\alpha)}$ to be non-degenerate. This
assumption is justified if we restrict $\omega$ to
$C_{0}^{(\alpha)}\times \mathbf{k}\subset \mathcal{U}_{\alpha}$.
We will have more to say about this restriction at the end of this section and also in Sec. 6.

Now, thanks to the condition $\pounds_{e_{c}}\omega_{ab}-\pounds_{e_{a}}\omega_{cb}-f_{ca}^{d}\omega_{db}=0$ and the fact that
$\frac{\partial}{\partial p^{c}}\omega_{ab}=0$,  we have
\begin{eqnarray}
d\left.
\omega\right\vert_{\mathcal{U}_{\alpha}}&=&\pounds_{e_{c}}\omega^{(\alpha)}_{ab}c^{c}\wedge
\overline{c}^{a}\wedge c^{b}-\omega^{(\alpha)}_{ad}
\overline{c}^{a}\wedge dc^{d}\nonumber\\
&=&-\pounds_{e_{c}}\omega^{(\alpha)}_{ab} \overline{c}^{a}\wedge
c^{c}\wedge
c^{b}+\frac{1}{2}f^{d}_{cb}\omega^{(\alpha)}_{ad}\overline{c}^{a}\wedge
c^{c}\wedge c^{b}\nonumber\\
&=&-\frac{1}{2}\left[\pounds_{e_{c}}\omega^{(\alpha)}_{ab}-\pounds_{e_{b}}\omega^{(\alpha)}_{ac}
-f^{d}_{cb}\omega^{(\alpha)}_{ad}\right]\overline{c}^{a}\wedge
c^{c}\wedge c^{b}=0.
\end{eqnarray}
Here we also made use of the Mauer-Cartan equation
$dc^{d}=-\frac{1}{2}f^{d}_{cb}c^{c}\wedge c^{b}$. Thus each
$C_{0}^{(\alpha)}\times \mathbf{k}$ becomes a symplectic manifold with symplectic
form $\left. \omega\right\vert_{C_{0}^{(\alpha)}\times \mathbf{k}}$. On each
$C_{0}^{(\alpha)}\times \mathbf{k}$ the action of $\mathbf{k}$ is free and
symplectic. The first assertion is true since the action of
$\mathbf{k}$ on itself and hence on
$C_{0}^{(\alpha)}\times \mathbf{k}$ is
free. The second assertion follows from a simple computation
\begin{eqnarray}
\pounds_{a}\omega^{(\alpha)}=\omega_{bc}(q) \pounds_{a}\overline{c}^{b}\wedge c^{c}=0
\end{eqnarray}
since $\pounds_{a}\overline{c}^{b}=\pounds_{a}dp^{b}=d\delta_{a}^{b}=0$.
 We can also show that this action is Hamiltonian by noticing that the equation defining the moment map
\begin{equation}
d\mu_{c}=-\iota_{c}\omega
\end{equation}
is equivalent to
\begin{eqnarray}{\label{p}}
  \pounds_{e_{b}} \mu_{a}= -\omega_{ab},\;\;\;\;\frac{\partial \mu_{a}}{\partial \pi^{b}}= 0.
\end{eqnarray}
Here the second equation implies that the moment map is independent of fiber coordinates. Then the first equation can be integrated to give the moment map as a function of $q$'s. Moreover
\begin{equation}
\left\lbrace \mu_{a},\mu_{b}\right\rbrace=\omega\left( \frac{\partial}{\partial \pi^{a}},\frac{\partial}{\partial \pi^{b}}\right) =0.
\end{equation}
Hence we conclude that the action is Hamiltonian. Thus we have proved
\begin{sub}
For each $\alpha$ the action of the additive group $\mathbf{k}$ on $C_{0}^{(\alpha)}\times \mathbf{k}$ is free, symplectic and Hamiltonian.
\end{sub}

One can form the Cartan model of equivariant cohomology based on
the action of the additive group $\mathbf{k}$ on $C_{0}^{(\alpha)}\times \mathbf{k}$ (in fact, also on $\mathcal{U}_{\alpha}$).
\begin{eqnarray}
C_{\mathbf{k}}(\Omega(C_{0}^{(\alpha)}\times \mathbf{k}))&=&(\mathcal{S(\mathbf{k}^{*})}\otimes \Omega(C_{0}^{(\alpha)}\times \mathbf{k}))^{\mathbf{k}}\\
&=& S(\mathbf{k}^{*})\otimes \left( \Omega(C_{0}^{(\alpha)}\times \mathbf{k})\right) ^{\mathbf{k}}.
\end{eqnarray}
Thus the generators of the Cartan model consist of $q$, $c$, $\overline{c}$ and $b$. In particular the tensor components of the differential form part of an element of Cartan model should be independent of $\pi$'s.
This model for the equivariant cohomology is a geometric version of the algebraic model
constructed at the beginning of this section by adding a trivial pair to the minimal sector of the BRST complex.
In fact the action of the Cartan derivative
\begin{equation}
D=d-ib^{a}\iota_{a}
\end{equation}
on the generators of the Cartan model is given by
\begin{eqnarray}
Dq^{a}&=&dq^{a}=c^{b}\pounds_{e_{b}}q^{a}\\
Db^{a}&=&0\\
D\overline{c}^{a}&=&-ib^{c}\iota_{c}dp^{a}=-ib^{a}\\
Dc^{a}&=&dc^{a}=-\frac{1}{2}f_{bc}^{a}\;c^{b}\wedge c^{c}
\end{eqnarray}
where we used the Mauer-Cartan equations in the last line.

Now using this relation between BRST cohomology and equivariant
cohomology, and the fact that for free group actions the latter is
equivalent to the de Rham cohomology of the quotient we get (for the
standard treatment see \cite{ht})
\begin{sub}
  \begin{equation}
H^{i}_{BRST}\cong H^{i}_{dR}(K), \;\;\; i\geq 0
\end{equation}
\end{sub}
\textit{Proof:} Notice that both the BRST cohomology and the equivariant cohomology contain the trivial pair $(b,\overline{c})$ which does not affect the cohomology of the minimal BRST
sector. Thus
\begin{equation}
H^{i}_{Min. \,BRST}\cong H^{i}_{Nonmin. \,BRST}\cong H^{i}_{\mathbf{k}}(\mathcal{U}_{\alpha}).
\end{equation}
But recall that the action of $\mathbf{k}$ on $\mathcal{U}_{\alpha}$ is free. Therefore using the fundamental characterization of the equivariant cohomology we have
\begin{equation}
H^{i}_{\mathbf{k}}(\mathcal{U}_{\alpha})\cong H^{i}_{dR}(\mathcal{U}_{\alpha}/\mathbf{k})\cong H^{i}_{dR}( \mathcal{O}_{\alpha})\cong H^{i}_{dR}(K).
\end{equation}
Thus we conclude
\begin{equation}
H^{i}_{BRST}\cong H^{i}_{dR}(K).
\end{equation}

Before we end this section we want to take a closer look at the singularities of $\omega$. 
We will denote the image of $\mu^{-1}(0)\subset \mathcal{O}_{\alpha}\times \mathbf{k}$ under the natural projection on $\mathcal{O}_{\alpha}$ by $\widetilde{\mu}^{-1}(0)$. 
Consider $\ell_{r}^{(\alpha)}$ and let $\left\lbrace X_{k}^{b} \right\rbrace_{k=1}^{r}$ be the zero modes of $\omega_{ab}$ i.e. $\omega_{ab}X_{k}^{b}=0$. Let $X_{k}=X_{k}^{b}e_{b}$ then 
\begin{equation}
\iota_{X_{k}}\omega=-\omega_{ab}\overline{c}^{a}X_{k}^{b}=0
\end{equation}
and 
\begin{equation}
\pounds_{X_{k}}\omega=d\iota_{X_{k}}\omega=0. 
\end{equation} 
Consequently $\pounds_{X_{k}}\omega^{n}=0$. On the other hand
\begin{eqnarray}
\pounds_{X_{k}}\omega^{n}&=&\pm\pounds_{X_{k}}\left[(\mathrm{det}\omega)\overline{c}^{1}\wedge\ldots\wedge\overline{c}^{1}\wedge c^{1}\wedge\ldots\wedge c^{n} \right]=\nonumber\\
&=&\pm\left[\pounds_{X_{k}}(\mathrm{det}\omega)\right] \overline{c}^{1}\wedge\ldots\wedge\overline{c}^{1}\wedge c^{1}\wedge\ldots\wedge c^{n}+\nonumber\\
&&\pm\sum_{a=1}^{n} (\mathrm{det}\omega)\overline{c}^{1}\wedge\ldots\wedge\overline{c}^{n}\wedge c^{1}\wedge\ldots\pounds_{X_{k}}c^{a}\ldots\wedge c^{n}=\nonumber\\
&=&\pm\left[\pounds_{X_{k}}(\mathrm{det}\omega)+e_{a}X^{a}_{k}(\mathrm{det}\omega) \right]\overline{c}^{1}\wedge\ldots\wedge\overline{c}^{n}\wedge c^{1}\wedge\ldots\wedge c^{n}\nonumber\\
&=&\pm\left[\pounds_{X_{k}}(\mathrm{det}\omega) \right]\overline{c}^{1}\wedge\ldots\wedge\overline{c}^{n}\wedge c^{1}\wedge\ldots\wedge c^{n}  
\end{eqnarray} 
where we used
\begin{eqnarray}
\pounds_{X_{k}}c^{a}&=&(\iota_{X_{k}}d+d\iota_{X_{k}})c^{a}\nonumber\\
&=&-\frac{1}{2}\iota_{X_{k}}(f_{bc}^{a}c^{b}\wedge c^{c})+dX^{a}\nonumber\\
&=&-f^{a}_{bc}X^{b}c^{c}+c^{b}e_{b}(X^{a})
\end{eqnarray} 
which, together with the complete anti-symmetry of $f^{a}_{bc}$, implies
\begin{eqnarray}
\overline{c}^{1}\ldots\overline{c}^{n}\wedge c^{1}\wedge\ldots\pounds_{X_{k}}c^{a}\ldots\wedge c^{n}=e_{a}(X^{a})\overline{c}^{1}\wedge\ldots\wedge\overline{c}^{1}\wedge c^{1}\wedge\ldots\wedge c^{n} 
\end{eqnarray} 
So we have $\pounds_{X_{k}}(\mathrm{det}\omega)=0$ and consequently $X_{k}$ is tangent to $\ell_{r}^{(\alpha)}$.
Moreover,
\begin{eqnarray}
\pounds_{\left[ X_{k},X_{l}\right] }\omega&=&d\iota_{\left[ X_{k},X_{l}\right] }\omega\nonumber\\
&=&d(\pounds_{X_{k}}\iota_{X_{l}}-\iota_{X_{l}}\pounds_{X_{k}})\omega=0
\end{eqnarray} 
which implies $\omega_{ab}\left[X_{k},X_{l} \right]^{b}=0 $. So $\left[X_{k},X_{l} \right]^{b}=F_{kl}^{m}X_{m}^{b}$ and $\left[X_{k},X_{l} \right]=F_{kl}^{m}X_{m}$. Thus the vector fields
$\left\lbrace X_{k} \right\rbrace_{k=1}^{r}$ form an integrable distribution on $\ell_{r}^{(\alpha)}$.
On each leaf of the corresponding foliation the moment map is constant:
\begin{equation}{\label{ma}}
\pounds_{X_{k}}\mu_{a}=X^{b}_{k}\pounds_{e_{b}}\mu_{a}=-\omega_{ab}X^{b}_{k}=0.
\end{equation} 
So we conclude that if $\widetilde{\mu}^{-1}(0)$ intersects $\ell_{r}^{(\alpha)}$ then the points of intersection form
a submanifold (union of the leaves with $\mu=0$) of the horizon.

Conversely, if a subset $\mathcal{N}$ of $\widetilde{\mu}^{-1}(0)$ is a smooth connected manifold in $\mathcal{O}_{\alpha}$ then $\mathcal{N}$ should lie on a horizon. If  $X$ is a vector field tangent to $\mathcal{N}$ then $\pounds_{X}\mu_{a}=0$ implies, by a calculation similar to (\ref{ma}) \cite{gribov}, $\omega_{ab}X^{b}=0$. Thus $X$ must be tangent to a horizon. We conclude that $\mathcal{N}$ lies on a horizon.

Summarizing, we have proved
\begin{sub}
If $\widetilde{\mu}^{-1}(0)$ intersects a horizon then it does so at certain leaves of the foliation of the horizon generated by the zero modes of $\omega_{ab}$. 
Conversely, if a subset of $\widetilde{\mu}^{-1}(0)$ forms a submanifold of $\mathcal{O}_{\alpha}$ then that submanifold lies on a horizon.
\end{sub}

\section{Generalization to Yang-Mills Theory}

As we remarked earlier in Yang-Mills theory $Q$ will be the space of
all connections on a principal $G$-bundle over the Euclidean space
$\mathbf{R}^{n}$. As is well known this is an affine space. We will
assume that $G$ is compact and its Lie algebra $\mathbf{g}$ is
semi-simple. The gauge group $K$ will be identified with the space
of all smooth maps from $\mathbf{R}^{n}$ into $G$. We will interpret
the ghost operator as a 2-form on $Q\times \mathbf{k}$.
The ghost fields $c^{a}(x)$ will be interpreted as 1-forms dual to the fundamental vector fields that generate infinitesimal gauge transformations.
\begin{eqnarray}
  e_{ax} = \int\,d^{n}y\,\left[\frac{\partial}{\partial
  y^{\mu}}\delta(x-y)\delta^{b}_{a}+f^{b}_{da}A_{\mu}^{d}(y)\delta(x-y)\right]\frac{\delta}{\delta A^{
  b}_{\mu}(y)}
\end{eqnarray}
\begin{eqnarray}
  \left[e_{ax},e_{by}\right]=\int d^{n}z\,f^{cz}_{ax,by}e_{cz}.
\end{eqnarray}
where $f^{cz}_{ax,by}=f^{c}_{ab}\delta(x-z)\delta(x-y)$ are the structure constants of the gauge algebra.

In order to generalize the construction of the last section to the case of Yang-Mills theory we have to
specify what is the field that generalizes fiber coordinates. Then the differential of this field can be
identified with the anti-ghost. In particular the field corresponding to fiber coordinates must be bosonic,
so that its differential will be anti-commuting, and must have the same index structure as the anti-ghost.
In the standard gauge fixed Yang-Mills action there is one such field, namely the auxiliary field $b^{a}(x)$.
However $b$ will be used as the generator of $\mathcal{S}(\mathbf{k}^{*})$, and therefore is not the correct
choice. Instead we introduce a non-interacting free field $\pi^{a}(x)$ into the Yang-Mills action. Such a
modification of the action will clearly not effect the value of the partition function, since $\pi$ is
Gaussian and do not interact with other fields. The introduction of $\pi$ as a free field into the Yang-Mills action allows us to
interpret the functional integral of the theory as an equivariant integral i.e. as an integral of a differential
form followed by an integral over $\mathcal{S}(\mathbf{k}^{*})$. More precisely, consider the partition function of the gauge fixed theory
\begin{equation}
Z=\int \mathcal{D}A \;\mathcal{D}b\; \mathcal{D}c\;
\mathcal{D}\overline{c}\; \mathcal{D}\pi\;e^{-S_{YM}-S_{gf}}R(\pi).
\end{equation}
Here $R(\pi)$ is the Gaussian regulator for integration over $\pi$; since neither $S_{YM}$ nor $S_{gf}$ depend on $\pi$ choosing the regulator such that $\int\mathcal{D}\pi\,R(\pi)=1$ gives us the usual partition function for the gauge fixed Yang-Mills action.
Assume that the gauge symmetry is not anomalous and write the measure for the gauge field as
\begin{equation}
\mathcal{D}A=J(\alpha)\;d\mu_{\alpha}\;\mathcal{D}U
\end{equation}
where $J(\alpha)$ is the gauge invariant Jacobian factor, $J(\alpha)d\mu_{\alpha}$ is the measure on the space of gauge orbits, and $\mathcal{D}U$ is the gauge invariant measure on the gauge orbit. Then the expectation value of a product $W$ of Wilson loops can be written as
\begin{equation}
\int_{M/K}
J(\alpha)\;d\mu_{\alpha}\;W\,e^{-S_{YM}}\int_{\alpha}
\mathcal{D}U \;\mathcal{D}b\; \mathcal{D}c\; \mathcal{D}\overline{c}\;
\mathcal{D}\pi\;e^{-S_{gf}}R(\pi).
\end{equation}
Now the integral
\begin{equation}
\int_{\alpha} \mathcal{D}U\; \mathcal{D}b\; \mathcal{D}c\;
\mathcal{D}\;\overline{c} \;\mathcal{D}\pi\;e^{-S_{gf}}R(\pi)
\end{equation}
is the integral of a differential form on $\mathcal{U}_{\alpha}=\mathcal{O}_{\alpha} \times \mathbf{k}$, followed by an integration over $\mathcal{S}(\mathbf{k}^{*})$. In fact the measures $\mathcal{D}U\;  \mathcal{D}c$ and
$\mathcal{D}\pi\;\mathcal{D}\overline{c} $ are the ones appropriate for integration of exterior forms on $\mathcal{O_{\alpha}}$ and $\mathbf{k}$, respectively.
\begin{equation}
\int_{\alpha} \mathcal{D}U\; \mathcal{D}c\;
\mathcal{D}\;\overline{c} \;\mathcal{D}\pi \rightarrow \int_{\mathcal{U}_{\alpha} =\mathcal{O}_{\alpha} \times \mathbf{k}}
\end{equation}

Now we want show that $S_{gf}$ is an equivariant form on $\mathcal{U}_{\alpha}$. We start by showing that
the ghost operator is a closed 2-form on $\mathcal{U}_{\alpha}$. We just need to check
the validity of the condition (\ref{q}) for the ghost operator $\partial^{\mu}\nabla_{\mu}$ whose integral kernel is
\begin{eqnarray}
  \omega_{ax,by} = \delta_{ab}\Box_{y}\delta(y-x)+\delta(y-x)(\partial\cdot
  A_{ab})(x)-
  A_{ab}(x)\cdot\partial_{y}\delta(y-x).
\end{eqnarray}
Thus we have to show
\begin{equation}
e_{cz}(\omega_{ax,by})-e_{by}(\omega_{ax,cz})=f^{du}_{cz,by}\omega_{ax,du}.
\end{equation}
The calculations, which are straightforward but a bit tedious, are given in the Appendix. Notice that
the above condition means that for each $(ax)$, $\omega_{ax,by}$ is a 1-cocycle in the cohomology of the gauge Lie algebra.

Next, we want show that $-\partial\cdot A_{a}(x)$ is the moment map for fiber translations. We need to check (\ref{p}) for $\mu_{ax}=-\partial\cdot A_{a}(x)$
\begin{eqnarray}
e_{yb}(-\partial\cdot A_{a}(x))&=&-\partial_{x}^{\mu}\left(\delta_{ab}\partial_{x \mu}\delta(y-x)+f_{adb}A^{d}_{\mu}(x)\delta(y-x)\right)\nonumber\\
&=&-\Box_{y}\delta(y-x)-(\partial\cdot A_{ab}(x)\delta(y-x)-A_{ab}\cdot\partial_{y}\delta(y-x)) \nonumber \\
&=&-\omega_{ax, by}
\end{eqnarray}
and
\begin{eqnarray}
\frac{\delta}{\delta \pi^{b}(y)}(-\partial\cdot A_{a}(x))=0.
 \end{eqnarray}
Thus the negative of the action for the ghost fields is an equivariant symplectic form
\begin{equation}
-S_{ghost}=\int d^{n}x \; \left[\overline{c}^{a}\partial^{\nu}\nabla_{\nu}c^{a}+ib^{a}\partial^{\nu}A_{\nu}^{a} \right]=\omega-ib\cdot \mu.
\end{equation}
where $b\cdot \mu$ is short for $\int d^{n}x \;b^{a}_{x}\mu_{ax}$.

More generally, the gauge fixing action is an equivariant form
\begin{equation}
-S_{gf}=-\frac{\epsilon}{2}b\cdot b+\omega-ib\cdot \mu=-\frac{\epsilon}{2}b\cdot b+\overline{\omega}
\end{equation}
So the integral
\begin{equation}
\int \mathcal{D}b\,\int_{\mathcal{U}_{\alpha}}  R(\pi) e^{-S_{gf}}=\int \mathcal{D}b\,\int_{\mathcal{U}_{\alpha}} R(\pi) e^{-\frac{\epsilon}{2}b\cdot b+\overline{\omega}}
\end{equation}
is almost the integral of an equivariant form. The problem is the Gaussian regulator $R$, which is not an equivariant form.  Moreover, we will have to address the fact that $\omega$ is degenerate on the Gribov horizons. In the next section we will see how one can handle these issues.

\section{Gauge Fixing and Equivariant Localization}

Now we are ready to interpret Faddeev-Popov method as equivariant localization. 
\subsection{Treatment of the Regulator}
The complication due to the regulator can be handled by incorporating the latter in the definition of equivariant integration. However it is important for our purposes to make sure that this modification does not spoil the equivariant localization principle.
\begin{sub}
For any equivariant form $\alpha$ in our Cartan model
\begin{equation}
\int R(\pi)D\alpha=0
\end{equation}
\end{sub}

\textit{Proof:} Let us write $\alpha$ as
\begin{equation}
    \alpha=\sum_{I}\alpha_{I}(q)b^{I}
\end{equation}
where
\begin{equation}
\alpha_{I}(q)=(\alpha_{I})_{KL}(q)dq^{K}d\pi^{L}.
\end{equation}
and $I$, $K$ and $L$ are multi-indices with a fixed value of $|K|+|L|$. Since $(\alpha_{I})_{KL}$ does not depend on
$p$ we have
\begin{equation}
d[(\alpha_{I})_{KL}dq^{K}d\pi^{L}]=d_{q}[(\alpha_{I})_{KL}dq^{K}]d\pi^{L}.
\end{equation}
So
\begin{eqnarray}
    \int
    R(\pi)D\alpha&=&\int
    R(\pi)d_{q}[(\alpha_{I})_{KL}dq^{K}d\pi^{L}]b^{I}\nonumber\\
    &&-R(\pi)(\alpha_{I})_{KL}\iota_{a}(dq^{K}d\pi^{L})b^{a}b^{I}\nonumber\\
    &=&\int
    R(\pi)d_{q}[(\alpha_{I})_{KL}dq^{K}d\pi^{L}]b^{I}\nonumber\\
    &=&\int_{\mathbf{k}} e^{-\epsilon(b,b)}b^{I}\;\int_{M}d_{q}[(\alpha_{I})_{KL}dq^{K}]\;\int_{\mathbf{k}}
    R(\pi)d\pi^{L}\nonumber\\
    &=&0.
\end{eqnarray}
As a simple corollary to this proposition we have the modified localization formula
\begin{sub}
  \begin{equation}
    \int R(\pi)\alpha= \int R(\pi)\alpha \;e^{tD\lambda}
   \end{equation}
for $D\alpha=0$  and any equivariant 1-form $\lambda$ in our Cartan model.
\end{sub}

\textit{Proof:}
\begin{eqnarray}
\int R(\pi)\;\alpha\;(e^{tD\lambda}-1)&=&\int R(\pi)\;\alpha \left(tD\lambda+\frac{t^{2}}{2}D\lambda D\lambda+... \right)\nonumber\\
&=&\int R(\pi)\;\alpha D\left(t\lambda+\frac{t^{2}}{2}\lambda D\lambda+...  \right)\nonumber\\
&=&\int R(\pi)\;D\left[ \alpha \left(t\lambda+\frac{t^{2}}{2}\lambda D\lambda+...  \right)\right]\nonumber\\
&=&0
\end{eqnarray}

\subsection{Choice of $J$ and $\lambda$}
The 1-form $\lambda$ will be chosen in the form $J(dI)$ where $J$ is a $\mathbf{k}$-invariant almost complex structure compatible with $\omega^{(\alpha)}$ and
\begin{equation}
I=\sum_{a}\mu_{a}^{2}.
\end{equation}
Let us assume for awhile that $\omega^{(\alpha)}$ is non-degenerate. We will consider the degenerate case later in this section.
As an invariant almost complex structure on $\mathcal{U}_{\alpha}$ we will take
 \begin{eqnarray}
J(c^{a})=-\frac{1}{2}(\omega^{-1})^{ab}\;\overline{c}^{b}\\
J(\overline{c}^{c})=2\omega^{cd}\;c^{d}
\end{eqnarray}
where all the indices are raised and lowered by $\delta$.
Thus
\begin{equation}
J=\frac{1}{2}(\omega^{-1})^{ab}\;\overline{c}^{b}\otimes e_{a}-2\omega^{ab}\;c^{b}\otimes \frac{\partial}{\partial \pi^{a}}
\end{equation}
\begin{sub}
$J$ is $\mathbf{k}$-invariant and compatible with $\omega$
\end{sub}
\textit{Proof:}
$\mathbf{k}$-invariance of $J$ follows from the fact that $\pounds_{a}$ annihilates all the terms appearing in $J$. $J$ is an almost complex structure as can be seen from:
\begin{eqnarray}
J^{2}(c^{a})&=&J\left( -\frac{1}{2}(\omega^{-1})^{ab}\;\overline{c}^{b}\right) =-(\omega^{-1})^{ab}\omega^{bc}c^{c}=-c^{a}\\
J^{2}(\overline{c}^{a})&=&J(2\omega^{ab}\;c^{b})=-\omega^{ab}(\omega^{-1})^{bc}\overline{c}^{c}=-\overline{c}^{a}.
\end{eqnarray}
So $J^{2}=-1$.
Compatibility of $J$ with $\omega$ follows from
\begin{eqnarray}
\omega_{ab}(J(\overline{c}^{a})\wedge J(c^{b}))&=&-\omega_{ab}\;(2\omega^{ac}c^{c})\wedge\frac{1}{2}(\omega^{-1})^{bd}\overline{c}^{d}\nonumber\\
&=&-\omega_{ab}(\omega^{-1})^{bd}\omega^{ac}c^{c}\wedge\overline{c}^{d}\nonumber\\
&=& \omega_{dc}\;\overline{c}^{d}\wedge c^{c}=\omega
\end{eqnarray}
and the fact that
\begin{eqnarray}
&&\omega_{ab}\;(J\overline{c}^{a})\otimes c^{b}-\omega_{ab}\;(Jc^{b})\otimes \overline{c}^{a}=\nonumber\\
&=&\omega_{ab}\left(2\omega^{ac}c^{c}\otimes c^{b}+\frac{1}{2}(\omega^{-1})^{bc}\overline{c}^{c}\otimes \overline{c}^{a}\right)\nonumber\\
&=&2(\omega^{T}\omega)_{cb}c^{c}\otimes c^{b}+\frac{1}{2}\overline{c}^{a}\otimes \overline{c}^{a}
\end{eqnarray}
is a positive definite metric.

The invariant 1-form $\lambda$ is given by
\begin{eqnarray}
\lambda&=&J(dI)=2J\left( \mu_{b}\;d\mu_{b}\right) =2J\left( \mu_{b}(\pounds_{e_{a}} \mu_{b}) c^{a}\right) =\nonumber\\
&=&-2\mu_{b}\;(\pounds_{e_{a}} \mu_{b})\;\frac{1}{2}(\omega^{-1})^{ad}\overline{c}^{d}=\mu_{b}\;\omega_{ba}(\omega^{-1})^{ad}\overline{c}^{d}=\nonumber\\
&=& \mu_{b}\overline{c}^{b}
\end{eqnarray}
and
\begin{eqnarray}
\iota_{b}\lambda=\sum_{c}\mu_{c}\;dp^{c}(\frac{\partial}{\partial p^{b}})=\mu_{b}
\end{eqnarray}

Let us define the vector field $U=-\mu_{c}(\omega^{-1})^{bc}e_{b}$ then
\begin{sub}
\begin{equation}
\lambda=\iota_{U}\omega,\;\;\;\;d\lambda=\omega,\;\;\mathrm{and}\;\;D\lambda=\overline{\omega}
\end{equation}
\end{sub}
\textit{Proof:}
\begin{eqnarray}
\iota_{U}\omega&=&\iota_{U}(\omega_{ad}\overline{c}^{a}\wedge c^{d})=\omega_{ad}\overline{c}^{a}\mu_{c}(\omega^{-1})^{bc}\delta^{d}_{b}\nonumber\\
&=&\mu_{a}\overline{c}^{a}=\lambda.
\end{eqnarray}
\begin{eqnarray}
d\lambda&=&d\mu_{b}\wedge \overline{c}^{b}=(-\iota_{b}\omega)\wedge \overline{c}^{b}\nonumber\\
&=&-\iota_{b}(\omega_{ac}\overline{c}^{a}\wedge \overline{c}^{c})\wedge \overline{c}^{b}\nonumber\\
&=&-\omega_{bc}c^{c}\wedge\overline{c}^{b}=\omega.
\end{eqnarray}
and
\begin{eqnarray}
D\lambda&=&(d-ib^{a}\iota_{a})\lambda=\omega-ib^{a}\mu_{a}=\overline{\omega}.
\end{eqnarray}

The following expression in terms of a coordinate basis will be useful later
\begin{eqnarray}
d\lambda&=&d\mu_{b}\wedge \overline{c}^{b}=\frac{\partial \mu_{b}}{\partial q^{i}}dq^{i}\wedge \overline{c}^{b}
\end{eqnarray}
The top form in $e^{td\lambda}$ is given by
\begin{equation}
\det\left(t \frac{\partial \mu_{b}}{\partial q^{i}}\right).
\end{equation}
The following is another useful identity
\begin{eqnarray}
\lambda \wedge\omega^{n-1} &=&\frac{1}{n}\;\iota_{U}\omega^{n} \nonumber\\
&=&\frac{1}{n}\;\iota_{U}\left[ \left(\omega_{a_{1}b_{1}}\ldots \omega_{a_{n}b_{n}}\right)\overline{c}^{a_{1}}\wedge c^{b_{1}}\wedge \ldots  \wedge \overline{c}^{a_{n}}\wedge c^{b_{n}}\right] \nonumber\\
&=& \frac{\varepsilon_{n}}{n}\;\iota_{U} \left[ \left(\omega_{a_{1}b_{1}}\ldots \omega_{a_{n}b_{n}}\right)\left( \overline{c}^{a_{1}}\wedge \ldots \wedge \overline{c}^{a_{n}}\right)\right.\wedge \left.\left(c^{b_{1}}\wedge \ldots \wedge c^{b_{n}}\right)\right] \nonumber  \\
&=& \frac{\varepsilon_{n}}{n}\;\iota_{U}\left[\omega_{a_{1}b_{1}}\ldots \omega_{a_{n}b_{n}}\epsilon^{a_{1}\ldots a_{n}} \epsilon^{b_{1}\ldots b_{n}}\overline{c}^{1}\wedge \ldots  \wedge \overline{c}^{n}\wedge c^{1}\wedge \ldots \wedge c^{n}\right] \nonumber\\
&=&\sigma_{n}(n-1)!\, (\mathrm{det} \omega) (\overline{c}^{1}\wedge \ldots \wedge \overline{c}^{n})\wedge \iota_{U} (c^{1}\wedge \ldots \wedge c^{n})
\end{eqnarray}
Here and in the following $\varepsilon_{n}=(-1)^{\frac{n(n-1)}{2}}$ and $\sigma_{n}=(-1)^{\frac{n(n+1)}{2}}$.
Notice that the almost complex structure $J$ and the metric $g$ are singular at the points where  $\omega$ is singular. Now, restricting all these structures to $C_{0}^{(\alpha)}\times \mathbf{k}$, where $\omega$ is non-degenerate, we get a positive definite metric $g$, which is devoid of singularities.
\subsection{Faddeev-Popov as Equivariant Localization}
Since $g$ has no singularities on $C_{0}^{(\alpha)}\times \mathbf{k}$, the equations
\begin{eqnarray}
b^{a}d\left[\iota_{a}\lambda \right]&=&0\\
 \iota_{a}\lambda=\mu_{a}&=&0 \;\;\forall a
\end{eqnarray}
that define the localizing manifold enjoy the nice properties discussed
in Sec. 3. In particular, the unique solution of the first equation
on $\mu_{0}^{-1}(0)=\mu^{-1}(0)\cap (C_{0}^{(\alpha)}\times \mathbf{k})$ is $b=0$.
Moreover, from the second equation we see that higher critical points do not contribute to localization. Below we will show that the contribution of $\mu^{-1}(0)$ is gauge independent (i.e. independent of $\epsilon$).
Now we must show that equivariant localization principle holds on
$C_{0}^{(\alpha)}\times \mathbf{k}$, that is, we must show that the
difference
\begin{eqnarray}
\Delta&\equiv&\int d^{n}b \int_{C_{0}^{(\alpha)}\times \mathbf{k}}
R(\pi)\;e^{-\frac{\epsilon}{2} b\cdot
b}e^{\overline{\omega}}\;(e^{tD\lambda}-1)
\end{eqnarray}
vanishes. As we will see shortly, this is proportional to a surface
term on $\ell_{1}^{(\alpha)}$. But before we do that let us try a
simpler localization, still based on equivariant cohomology, to
illustrate the problems encountered in restricting the path integral
to $C_{0}^{(\alpha)}\times \mathbf{k}$.

Let us note that
\begin{equation}
-\frac{1}{2}\epsilon b\cdot b=D\beta
\end{equation}
where $\beta$ is the equivariant 3-form
\begin{eqnarray}
\beta=-\frac{i}{2}\epsilon b_{a}\overline{c}^{a},\;\;\;d\beta=0.
\end{eqnarray}
So one may write the partition function as
\begin{eqnarray}
\int_{C_{0}^{(\alpha)}\times
\mathbf{k}}e^{\overline{\omega}}R(\pi)=\int
d^{n}b\int_{C_{0}^{(\alpha)}\times
\mathbf{k}}R(\pi)e^{D\beta}e^{-ib\cdot\mu+\omega}.
\end{eqnarray}
The term $e^{D\beta}$ can be omitted in this integral if one can show that
\begin{eqnarray}
\Delta'\equiv\int d^{n}b\int_{C_{0}^{(\alpha)}\times
\mathbf{k}}R(\pi)\left( e^{D\beta}-1\right)
e^{-ib\cdot\mu+\omega}=0.
\end{eqnarray}
If this was the case then one could integrate over $b$ to get a delta function and thus localize the integral. However it is not clear that the above difference is zero. The problem is a nonvanishing surface term.
\begin{eqnarray}
\Delta'&=&\int d^{n}b\int_{C_{0}^{(\alpha)}\times \mathbf{k}}D\left[\left(\beta+\frac{1}{2}\beta D\beta \ldots \right)  e^{\overline{\omega}}\right]R(\pi)\nonumber\\
&=& \int d^{n}b\int_{C_{0}^{(\alpha)}\times \mathbf{k}} d\left[\left(\beta+\frac{1}{2}\beta D\beta \ldots \right)  e^{\overline{\omega}}\right]R(\pi)\nonumber\\
&=& -\int d^{n}b\int_{C_{0}^{(\alpha)}\times \mathbf{k}}\beta(d  e^{\overline{\omega}}) F(b^{2})R(\pi)
\end{eqnarray}
where
\begin{equation}
F(b^{2})=1+\frac{1}{2}D\beta+\ldots
\end{equation}
is a power series in $b^{2}=b\cdot b$, and we used $D\overline{\omega}=0=d\beta$. Thus
\begin{eqnarray}
\Delta'&=&-\int d^{n}b\int_{C_{0}^{(\alpha)}\times \mathbf{k}}\beta d\left[e^{-ib\cdot \mu}e^{-\omega} \right] F(b^{2})R(\pi)\nonumber\\
&=& -\int d^{n}b\int_{C_{0}^{(\alpha)}\times \mathbf{k}}\frac{1}{(n-1)!}\beta d\left[e^{-ib\cdot \mu}\omega^{n-1} \right]F(b^{2})R(\pi)
\end{eqnarray}
noting
\begin{equation}
d\left[e^{-ib\cdot\mu}\omega^{n-1} \right]=\pm \overline {c}^{a_{1}}\wedge\ldots\wedge \overline{c}^{a_{n-1}}d\left[e^{-ib\cdot\mu}\omega_{a_{1}b_{1}}\ldots \omega_{a_{n-1}b_{n-1}} c^{b_{1}}\wedge\ldots\wedge c^{b_{n-1}}\right],
\end{equation}
integrating over the fiber $\mathbf{k}$, and using Stokes theorem
for the integral over $C_{0}^{(\alpha)}$
\begin{equation}
\Delta'=\pm\int d^{n}b \frac{ \epsilon
F(b^{2})}{2(n-1)!}b_{a}\int_{\ell_{1}^{(\alpha)}}\epsilon^{a a_{1}\ldots
a_{n-1}}e^{-ib\cdot\mu}\omega_{a_{1}b_{1}}\ldots
\omega_{a_{n-1}b_{n-1}}c^{b_{1}}\ldots c^{b_{n-1}}
\end{equation}
which is in general non-zero. As expected, $\Delta'=0$ is guaranteed only for the special case $\epsilon=0$ (Landau gauge).

Now let us go back and try to localize the integral by using the invariant 1-form $\lambda$. A consistent use of equivariant localization principle on $C_{0}^{(\alpha)}\times \mathbf{k}$ requires the vanishing of the difference
\begin{eqnarray}
\Delta&\equiv&\int \left[ d^{n}b\right]  \int_{C_{0}^{(\alpha)}\times \mathbf{k}} R(\pi)\;e^{\overline{\omega}}\;(e^{tD\lambda}-1)\\
&=&\int \left[ d^{n}b\right]  \int_{C_{0}^{(\alpha)}\times
\mathbf{k}}
R(\pi)\;e^{\overline{\omega}}\;(e^{t\overline{\omega}}-1).
\end{eqnarray}
Here $[d^{n}b]=d^{n}b\,e^{-\frac{1}{2}\epsilon b\cdot b}$. So
\begin{equation}
\Delta=0 \;\;\Leftrightarrow \;\; \int \left[ d^{n}b\right]
\int_{C_{0}^{(\alpha)}\times \mathbf{k}}
R(\pi)\;e^{(t+1)\overline{\omega}}=\int \left[ d^{n}b\right]
\int_{C_{0}^{(\alpha)}\times \mathbf{k}}
R(\pi)\;e^{\overline{\omega}}
\end{equation}
or
\begin{equation}
\Delta=0 \;\;\Leftrightarrow \;\; \frac{\partial}{\partial t}\int
\left[ d^{n}b\right]  \int_{C_{0}^{(\alpha)}\times \mathbf{k}}
R(\pi)\;e^{t\overline{\omega}}=0.
\end{equation}
This is very similar to the situation encountered in topological field theories where $\Delta \neq 0$ signals the brakedown of BRST symmetry \cite{witten6}.

\begin{eqnarray}
\frac{\partial}{\partial t}\int \left[ d^{n}b\right]  \int_{C_{0}^{(\alpha)}\times \mathbf{k}} R(\pi)\;e^{t\overline{\omega}}&=&\int \left[ d^{n}b\right]  \int_{C_{0}^{(\alpha)}\times \mathbf{k}} R(\pi)\;\overline{\omega}\;e^{t\overline{\omega}}\nonumber\\
&=&\int \left[ d^{n}b\right]  \int_{C_{0}^{(\alpha)}\times \mathbf{k}} R(\pi)\;D\lambda\;e^{t\overline{\omega}}\nonumber\\
&=&\int \left[ d^{n}b\right]  \int_{C_{0}^{(\alpha)}\times \mathbf{k}} R(\pi)\;D\left( \lambda\;e^{t\overline{\omega}}\right) \nonumber\\
&=&\int \left[ d^{n}b\right]  \int_{C_{0}^{(\alpha)}\times
\mathbf{k}} R(\pi)\;d\left( \lambda\;e^{t\overline{\omega}}\right).
\end{eqnarray}
The top exterior form in $d\left(
\lambda\;e^{t\overline{\omega}}\right)$ is
\begin{eqnarray}
\frac{t^{n-1}}{(n-1)!}d\,\left[e^{-it b\cdot\mu}\,\lambda \wedge \omega^{n-1}\right]&=&\sigma_{n}\,t^{n-1}d\left[ \,e^{-it b\cdot\mu} (\mathrm{det} \omega)(\overline{c}^{1}\wedge \ldots \wedge \overline{c}^{n})\right.   \nonumber\\
&&\left. \wedge \iota_{U} (c^{1}\wedge \ldots \wedge c^{n})\right]
\end{eqnarray}
After integrating over $\mathbf{k}$ we get
\begin{eqnarray}
&&\frac{\partial}{\partial t}\int \left[ d^{n}b\right]  \int_{C_{0}^{(\alpha)}\times \mathbf{k}} R(\pi)\;e^{t\overline{\omega}}=\nonumber\\
&&=\sigma_{n}t^{n-1}\,\int \left[ d^{n}b\right]
\int_{\ell_{1}^{(\alpha)}}e^{-it b\cdot\mu}
\det(\omega)\left[\iota_{U}(c^{1}\wedge \ldots \wedge c^{n})
\right]=0
\end{eqnarray}
since $\det(\omega)=0$ on $\ell_{1}^{(\alpha)}$. Thus we conclude $\Delta=0$ and the equivariant localization principle is valid on $C_{0}^{(\alpha)}\times \mathbf{k}$.

Now we can localize the equivariant integral on the critical point set $\mu^{-1}_{0}(0)$.
Assuming the gauge slice intersects $C_{0}^{(\alpha)}$ only once
we have $\mu^{-1}_{0}(0)=\left\lbrace q^{(\alpha)}\right\rbrace \times \mathbf{k} $ where $q^{(\alpha)}$ is the element of  $C_{0}^{(\alpha)}$ with $\mu(q^{(\alpha)})=0$; i.e. the point of intersection. We do not expect contributions from the boundary since the integrand is proportional to $\mathrm{det}\omega$.
In the equivariant integral one can now localize the integration over $C_{0}^{(\alpha)}\times \mathbf{k}$ on a neighborhood
of $\mu^{-1}_{0}(0)$ of the form $B \times \mathbf{k}$ where $B$ is a ball with center at $q^{(\alpha)}$. The integral over $b$ can also be localized on a neighborhood $S_{0}$ of $b=0$.
\begin{eqnarray}
\int d^{n}b \int_{C_{0}^{(\alpha)}\times \mathbf{k}} e^{-\frac{1}{2}\epsilon b\cdot b+\overline{\omega}}e^{tD\lambda}R(\pi)=\int_{S_{0}} d^{n}b \int_{B \times \mathbf{k}} e^{-\frac{1}{2}\epsilon b\cdot b+\overline{\omega}}e^{tD\lambda}R(\pi).
\end{eqnarray}
Since $\mathbf{k}$ acts freely on $B \times \mathbf{k}$ we get
\begin{eqnarray}
-\frac{\epsilon}{2} b\cdot b\in H^{4}_{\mathbf{k}}(B \times \mathbf{k})\simeq H^{4}((B \times \mathbf{k})/\mathbf{k})=H^{4}(B)=0.
\end{eqnarray}
In fact, as we saw earlier,
\begin{eqnarray}
-\frac{1}{2}\epsilon b\cdot b=D\beta, \;\;\;\beta=-\frac{i}{2}\epsilon b\cdot c.
\end{eqnarray}
Now we have the following (see also Sec. 2 of \cite{witten1})
\begin{sub}
\begin{eqnarray}
\int_{S_{0}} d^{n}b\int_{B \times \mathbf{k}} e^{D\beta}e^{(t+1)D\lambda}R(p)=\int_{S_{0}} d^{n}b\int_{B \times \mathbf{k}} e^{(t+1)D\lambda}R(\pi)
\end{eqnarray}
\end{sub}

\textit{Proof:} Define a $\mathbf{k}$ invariant (i.e. independent of $\pi$) bump function $u$ which is equal to 1 in a neighborhood $B'\times \mathbf{k}$, $B'\subset B$, of $q^{(\alpha)}$ and 0 outside $B \times \mathbf{k}$ then the partition function can be written as
\begin{eqnarray}
\int_{S_{0}} d^{n}b\int_{B \times \mathbf{k}} e^{D\beta}e^{(t+1)D\lambda}R(\pi)u(q).
\end{eqnarray}
Now consider the difference
\begin{eqnarray}
&&\int_{S_{0}} d^{n}b\int_{B \times \mathbf{k}} \left( e^{D\beta}-1\right) e^{(t+1)D\lambda}R(\pi)u(q)=\nonumber\\
&&\int_{S_{0}} d^{n}b\int_{B \times \mathbf{k}} D\left(\beta+\frac{1}{2}\beta D\beta+\ldots \right) e^{(t+1)D\lambda}R(\pi)u(q)=\nonumber\\
&&\int_{S_{0}} d^{n}b\int_{B \times \mathbf{k}} d\left[ \left(\beta+\frac{1}{2}\beta D\beta+\ldots\right)   e^{(t+1)D\lambda}u(q)\right] R(\pi)+\nonumber\\
&&+\int_{S_{0}} d^{n}b\int_{B \times \mathbf{k}} \left(\beta+\frac{1}{2}\beta D\beta+\ldots\right)   e^{(t+1)D\lambda}(du(q))R(\pi) 
\end{eqnarray}
Performing the integration over $\pi$ in the first integral gives a surface term proportional to $u$; therefore the first integral vanishes. On the other hand in the large $t$ limit the second integral can be restricted onto $B'\times \mathbf{k}$ where $du=0$; consequently the second integral vanishes as well \cite{witten1}.

After letting $t\rightarrow t-1$ we get
 \begin{eqnarray}
\int  \;e^{\overline{\omega}}e^{(t-1)D\lambda}R(\pi)&=&\int d^{n}b\int_{C_{0}^{(\alpha)}\times \mathbf{k}}\;e^{tD\lambda}R(\pi)\nonumber\\
&=& \int d^{n}b\int_{C_{0}^{(\alpha)}\times \mathbf{k}}\;e^{td\lambda-itb^{a}\iota_{a}\lambda}R(\pi)\nonumber\\
&=&\int_{C_{0}^{(\alpha)} \times \mathbf{k}}\delta^{(n)} \left(t \iota_{a}\lambda \right) \;\;e^{td\lambda}R(\pi)\nonumber\\
&=&\int_{C_{0}^{(\alpha)} \times \mathbf{k}}\delta^{(n)} \left(t \mu_{a} \right) \;\;e^{td\lambda}R(\pi)
\end{eqnarray}
but, as we noted earlier, the top form in $e^{ td\lambda}$ is
\begin{eqnarray}
\det\left[t \frac{\partial \mu_{a}}{\partial q^{i}} \right]
\end{eqnarray}
Therefore the integrand is
\begin{eqnarray}
\delta^{(n)}(t \mu_{a})\det\left[t \frac{\partial \mu_{a}}{\partial q^{i}} \right]   =\delta(q-q^{(\alpha)})\mathrm{sgn}\det\left[t \frac{\partial \mu_{a}}{\partial q^{i}} \right].
\end{eqnarray}

Using
\begin{eqnarray}
\frac{\partial \mu_{c}}{\partial q^{i}}=e_{i}^{b}\pounds_{e_{b}}\mu_{c}=e_{i}^{b}\omega_{bc}
\end{eqnarray}
and choosing $\det (e_{i}^{b})>0$ we get the integrand equal to
\begin{equation}
\delta(q-q^{\alpha})\; \mathrm{sgn} \det (\omega).
\end{equation}
But $\det(\omega)>0$ inside the first Gribov horizon. Therefore the integrand is simply a delta function around the point of intersection $q^{r}$ and consequently we get
\begin{equation}
\int R(p) \;e^{-\overline{\omega}}=1.
\end{equation}

If there are additional Gribov copies we obtain a sum over delta functions and the integral gives the intersection number $n_{\alpha}$ of the gauge slice with $C_{0}^{(\alpha)}$. In this case a furher truncation of the ordinary critical points may be necessary \cite{ zwanziger1, sts, antonio, parrinello, baal1}. We will not address this issue in this paper.

\section{Application to Topological Field Theories}
Our methods can readily be used to explain the localization
properties of certain topological field theories. Here, for
illustrative purposes, we will consider simplest such theory, namely
the supersymmetric quantum mechanics \cite{witten4, witten5}. 
\subsection{Supersymmetric Quantum Mechanics}
For simplicity we will consider the 1 dimensional case. According to Baulieu-Singer method \cite{bs} (see also \cite{bs2, brt}) the action of SUSY
quantum mechanics can be interpreted as the gauge fixing action
corresponding to the zero action on the space of histories of a
particle living on $ \mathbf{R}$. The zero action, being invariant
under an arbitrary shift
\begin{equation}
q(t)\rightarrow q(t)+\delta q(t),
\end{equation}
needs gauge fixing. Notice that the action of this abelian gauge group is transitive and that there is only one gauge orbit.
If the gauge condition/moment map is chosen as
\begin{equation}
\frac{dq}{dt}+\frac{\partial V}{\partial q}
\end{equation}
where $V$ is a polynomial, then the standard Faddeev-Popov procedure gives the gauge fixing action (which is equal to the total action) as
\begin{equation}
\int dt \left[ \frac{\epsilon}{2}b^{2}-ib\left(\frac{dq}{dt}+\frac{\partial V}{\partial q} \right)-\overline{\psi}\left(\frac{d}{dt}+\frac{\partial^{2}V}{\partial q^{2}} \right)\psi\right].
\end{equation}
We will assume that $V'$ and $V''$ do not have simultaneous zeros.
We will employ periodic boundary condition for $q(t)$ with period $T$:
\begin{equation}
q(0)=q(T).
\end{equation}

The action of BRST/Cartan differential on the generators is given by
\begin{eqnarray}
Dq&=&\psi\\
D\psi&=&0\\
D\overline{\psi}&=&-ib\\
Db&=&0.
\end{eqnarray}

First let us characterize the Gribov horizons of this action. The
Gribov horizon is defined as the locus where the pre-symplectic form
is degenerate. So we look for the zero modes $f$ of the ghost
operator
\begin{equation}
\frac{df}{dt}+\frac{\partial^{2}V}{\partial q^{2}}f=0.
\end{equation}
The solution as a functional of $q$ is given by
\begin{equation}
f(t)=f(0)\exp\left[-\int_{0}^{t}dt'\;V''(q(t')) \right].
\end{equation}
Since we employ periodic boundary conditions these zero modes will be present only when
 \begin{equation}
\int_{0}^{T}dt'\;V''(q(t'))=0.
\end{equation}
This condition specifies the Gribov horizons of our model. Thus in a
region lying between two consecutive Gribov horizons histories obey
\begin{equation}
\int_{0}^{T}dt'\;V''(q(t')) \neq 0.
\end{equation}
Notice that static configurations $q_{s}$ with $V''(q_{s})\neq 0$ do
not lie on Gribov horizons. Moreover, for certain potentials Gribov
horizons are absent. For example there are no Gribov horizons for a
potential $V(q)$ with $V''(q)> 0 \; \forall q$.

The ordinary critical points are given by the zeroes of the moment
map
\begin{equation}
\frac{dq}{dt}+\frac{\partial V}{\partial q}=0.
\end{equation}
These can be recognized as the instantons of the theory:
\begin{eqnarray}
I&=&\int_{0}^{T} dt\; \mu^{2}=\int_{0}^{T} dt \left( \frac{dq}{dt}\right)^{2}+ \left(\frac{\partial V}{\partial q} \right)^{2}+2 \left(\frac{dq}{dt}\frac{\partial V}{\partial q} \right)\\
&=&\int_{0}^{T} dt \left( \frac{dq}{dt}\right)^{2}+ \left(\frac{\partial V}{\partial q} \right)^{2}+2\left[V(q(T))-V(q(0)) \right]\\
&=& \int_{0}^{T} dt \left( \frac{dq}{dt}\right)^{2}+
\left(\frac{\partial V}{\partial q} \right)^{2}
\end{eqnarray}
but for an ordinary critical point we have
\begin{equation}
I= \int_{0}^{T} dt \left( \frac{dq}{dt}\right)^{2}+
\left(\frac{\partial V}{\partial q} \right)^{2}=0
\end{equation}
hence
\begin{eqnarray}
\frac{dq}{dt}&=&0 \\
V'(q)&=&0.
\end{eqnarray}
So the ordinary critical points are time independent and coincide
with the critical points of the potential $V$. Since $V'$ and $V''$
do not have common zeroes, we see that none of the critical points
lie on a Gribov horizon. Then the partition function is given by
\begin{equation}
    Z=\sum_{q_{c}}  \mathrm{sgn}
\det\left[\frac{d}{dt}+V''(q_{c}) \right].
\end{equation}
This is the result derived in \cite{bbrt} by Nicolai map construction \cite{nicolai}.
If we relax the assumption that $V'$ and $V''$ do not vanish simultaneously we get critical points that lie on Gribov horizons. In this case one can avoid the Gribov horizons by restricting the integral to regions lying between the horizons. Since the integrand is proportional to $\mathrm{det}\,\omega$ (after the anti-commuting fields are integrated out) the contribution of a horizon to $Z$ is zero.
The surface terms will also vanish since they are all proportional to
$\mathrm{det}\,\omega$ and we still get the same answer for the partition function.

If one relaxes the periodicity condition then one gets non-trivial
solutions of $\mu=0$ (instantons). These non-trivial instantons lie on Gribov
horizons (i.e. they have zero modes). Again their contributions to $Z$ vanish, since the integrand is proportional to $\mathrm{\det}\,\omega$. But for path integrals with BRST exact (i.e. equivariantly exact) insertions, the surface terms are no longer proportional to $\mathrm{det}\,\omega$. So, one cannot ensure the gauge independence of the equivariant integral and this signals the breakdown of the BRST symmetry/SUSY. In fact, it is well known that in this case the instanton effects break the supersymmetry \cite{sh, novikov, ouvry, bbrt}.

\paragraph{Acknowledgement:} The author would like to thank O. T. Turgut for useful conversations.

\appendix

\section {Closedness of the Ghost Operator}
\setcounter{equation}{0}
\renewcommand{\theequation}{\Alph{section}.\arabic{equation}}
 \renewcommand{\thesection}{\Alph{section}}
In this appendix we will show that the ghost operator
$\partial^{\mu}\nabla_{\mu}$ obeys the condition (\ref{q}) which implies
$\left.\omega\right|_{\mathcal{U}_{\alpha}}$ is a closed form. We
will assume that the space-time is the Euclidean space
$\mathbf{R}^{n}$. Then the gauge group $K$ can be identified as the
space of smooth maps from $\mathbf{R}^{n}$ into a compact Lie group
$G$. We will assume that the Lie algebra $\mathbf{g}$ of $G$ is
semi-simple. The structure constants $f^{c}_{ab}$ are antisymmetric
in the lower indices and obey the Jacobi identity
\begin{equation}
    f^{d}_{ab}f^{e}_{dc}+f^{d}_{ca}f^{e}_{db}+f^{d}_{bc}f^{e}_{da}=0
\end{equation}
We will take $\delta_{ab}$ as the Cartan-Killing form and use it to
raise and lower the Lie algebra indices. The fact that $G$ is
compact and $\mathbf{g}$ is semi-simple implies
$f_{cab}=\delta_{cd}f^{d}_{ab}$ is totally anti-symmetric.

In the adjoint representation the gauge connection is given by
\begin{equation}
    A_{\mu\,ab}=A_{\mu}^{c}f_{acb}.
\end{equation}
where $A_{\mu}^{c}$ is the connection in the fundamental
representation. As a consequence of the Jacobi identity we get
\begin{equation}{\label{J}}
    f_{dab}A^{\mu}_{dc}-f_{dac}A^{\mu}_{db}=f_{dbc}A^{\mu}_{ad}.
\end{equation}
The infinitesimal action of $K$ on $Q$ is given by the fundamental
vector fields
\begin{eqnarray}
  e_{ax} = \int\,d^{n}y\,\left[\frac{\partial}{\partial
  y^{\mu}}\delta(x-y)\delta^{b}_{a}+f^{b}_{da}A_{\mu}^{d}(y)\delta(x-y)\right]\frac{\delta}{\delta A^{
  b}_{\mu}(y)}
\end{eqnarray}
\begin{eqnarray}
  \left[e_{ax},e_{by}\right]=f^{c}_{ab}\delta(x-y)e_{cx}.
\end{eqnarray}
The structure constants of the gauge algebra are
\begin{eqnarray}
  f^{cz}_{ax,by}=f^{c}_{ab}\delta(x-y)\delta(x-z).
\end{eqnarray}
The ghost operator $\partial^{\mu}\nabla_{\mu}$ can be written
explicitly as
\begin{eqnarray}
\partial^{\mu}\nabla_{\mu
ab}=\partial^{\mu}\partial_{\mu}\delta_{ab}+(\partial^{\mu}A_{\mu
ab})+A_{\mu ab}\partial^{\mu}.
\end{eqnarray}
The integral kernel of this operator is
\begin{eqnarray}
  \omega_{ax,by} = \delta_{ab}\Box_{y}\delta(y-x)+\delta(y-x)(\partial\cdot
  A_{ab})(x)-
  A_{ab}(x)\cdot\partial_{y}\delta(y-x).
\end{eqnarray}
Now we want to verify that
\begin{equation}
e_{cz}(\omega_{ax,by})-e_{by}(\omega_{ax,cz})=f^{du}_{cz,by}\omega_{ax,du}.
\end{equation}
Let us start with the right hand side
\begin{eqnarray}
f^{du}_{cz,by}\omega_{ax,du}&=& \int d^{n}u \; f_{dcb}\delta(z-y)\delta(z-u)\omega_{ax,du}\nonumber \\
   &=&  f_{dcb}\delta(z-y)\omega_{ax,dz}\nonumber \\
   &=&  f_{dcb}\left[\delta_{ad}\Box_{z}\delta(z-x)+\delta(z-x)(\partial\cdot
  A_{ad})(x)\right.\nonumber\\
  &&\left. -A_{ad}(x)\cdot\partial_{z}\delta(z-x)\right]\nonumber\\
  &=&f_{acb}\delta(z-y)\Box_{z}\delta(z-x)+\delta(z-y)\delta(z-x)\partial\cdot( f_{dcb}A_{ad})(x)\nonumber\\
&&-f_{dcb}A_{ad}(x)\cdot\delta(z-y)\partial_{z}
  \delta(z-x)
\end{eqnarray}
On the other hand
\begin{eqnarray}
e_{cz}\omega_{ax,by}&=&\delta(y-x)\partial^{\mu}_{x}f_{aeb}\left[\partial_{x\mu}\delta(z-x)\delta^{e}_{c}+
A_{\mu ec}(z)\delta(z-x) \right]-\nonumber\\
&&f_{aeb}\left[\partial_{x\mu}\delta(z-x)\delta^{e}_{c}+
A_{\mu ec}(z)\delta(z-x) \right]\partial^{\mu}_{y}\delta(y-x)\nonumber\\
&=&f_{acb}\left[ \delta(y-x)\Box_{x}\delta(z-x)-\partial_{x\mu}\delta(z-x)\partial^{\mu}_{y}\delta(y-x)\right]+\nonumber\\
&&f_{aeb}A_{\mu ec}(z)\left[\delta(y-x)\partial_{x\mu}\delta(z-x)-\delta(z-x)\partial^{\mu}_{y}\delta(y-x) \right]\nonumber\\
\end{eqnarray}
Using the distributional identities:
\begin{eqnarray}
\left[\delta(y-x)\partial_{x\mu}\delta(z-x)-\delta(z-x)\partial^{\mu}_{y}\delta(y-x) \right]f_{\mu}(z)=\nonumber\\
\left[\delta(z-x)\partial_{x\mu}\delta(y-x)-\delta(y-x)\partial^{\mu}_{z}\delta(z-x) \right]f_{\mu}(z)=\nonumber\\
\delta(z-y)\delta(z-x)(\partial\cdot f)(x)-f_{\mu}(x)\delta(z-y)\partial^{\mu}_{z}\delta(z-x)
\end{eqnarray}
and
\begin{eqnarray}
\delta(y-x)\Box_{x}\delta(z-x)-\partial_{x\mu}\delta(z-x)\partial^{\mu}_{y}\delta(y-x)+(y\leftrightarrow z)=\nonumber\\
=\delta(z-y)\Box_{z}\delta(z-x)
\end{eqnarray}
we can write
\begin{eqnarray}
e_{cz}(\omega_{ax,by})-e_{by}(\omega_{ax,cz})&=&f_{acb}\delta(z-y)\Box_{z}\delta(z-x)+\nonumber\\
&&\delta(z-y)\delta(z-x)(f_{adb}\partial\cdot A_{dc}-f_{adc}\partial\cdot A_{db})+\nonumber\\
&&-(f_{adb}A_{\mu dc}-f_{adc}A_{\mu eb})\delta(z-y)\partial^{\mu}_{z}\delta(z-x).\nonumber\\
\end{eqnarray}
Finally using (\ref{J}) we get
\begin{eqnarray}
e_{cz}(\omega_{ax,by})-e_{by}(\omega_{ax,cz})&=&f_{acb}\delta(z-y)\Box_{z}\delta(z-x)+\nonumber\\
&&\delta(z-y)\delta(z-x)\partial\cdot(f_{dcb} A_{ad})+\nonumber\\
&&-(f_{dcb}A_{\mu ad})\delta(z-y)\partial^{\mu}_{z}\delta(z-x)\nonumber\\
\end{eqnarray}
which equals $f^{du}_{cz,by}\omega_{ax,du}$.

\end{document}